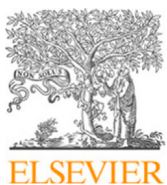

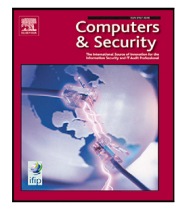

# Navigating quantum security risks in networked environments: A comprehensive study of quantum-safe network protocols

Yaser Baseri [a,*], Vikas Chouhan [b], Abdelhakim Hafid [a]

[a] Department of Computer Science and Operations Research, Universite de Montreal, Canada
[b] Canadian Institute for Cybersecurity (CIC), University of New Brunswick, Canada

## A R T I C L E  I N F O

## A B S T R A C T

The emergence of quantum computing poses a formidable security challenge to network protocols traditionally safeguarded by classical cryptographic algorithms. This paper provides an exhaustive analysis of vulnerabilities introduced by quantum computing in a diverse array of widely utilized security protocols across the layers of the TCP/IP model, including TLS, IPsec, SSH, PGP, and more. Our investigation focuses on precisely identifying vulnerabilities susceptible to exploitation by quantum adversaries at various migration stages for each protocol while also assessing the associated risks and consequences for secure communication. We delve deep into the impact of quantum computing on each protocol, emphasizing potential threats posed by quantum attacks and scrutinizing the effectiveness of post-quantum cryptographic solutions. Through carefully evaluating vulnerabilities and risks that network protocols face in the post-quantum era, this study provides invaluable insights to guide the development of appropriate countermeasures. Our findings contribute to a broader comprehension of quantum computing's influence on network security and offer practical guidance for protocol designers, implementers, and policymakers in addressing the challenges stemming from the advancement of quantum computing. This comprehensive study is a crucial step towards fortifying the security of networked environments in the quantum age.

## 1. Introduction

Quantum computing has emerged as a powerful field that combines principles from physics and computer science. It offers the potential to solve complex problems at an astonishing speed, revolutionizing various domains. However, this remarkable computational power raises concerns about the security of network protocols that rely on traditional cryptographic algorithms. Quantum computers have the ability to break widely-used encryption algorithms, such as RSA and Elliptic Curve Cryptography (ECC), which are the foundations of secure communication in network protocols (Shor, 2022; Grover, 2023). In organizations, security protocols are extensively employed to verify the source and protect the confidentiality and integrity of information during communication and storage. Current security protocols, such as TLS and SSH, are effective against classical computer attacks on network communications (Portmann and Renner, 2022). However, the emergence of fault-tolerant quantum computers could pose a significant threat to the security of these protocols by exploiting the underlying mathematical challenges in a matter of hours or even seconds (Preskill, 2018).

The growing impact of quantum computing on network protocols has recently drawn significant attention. As quantum computers continue to advance, their ability to breach traditional cryptographic algorithms raises serious concerns regarding network communication security. In response to this challenge, researchers have been actively exploring post-quantum cryptographic solutions capable of withstanding quantum computer attacks. However, the crucial step remains evaluating the specific impact of quantum computing on individual network protocols to assess the vulnerabilities they may face, ensuring the ongoing security of network communications. In light of this context, the focus of this work is dedicated to the examination of widely adopted standard network and security protocols, as referenced in the Canadian National Quantum-Readiness guidelines (Quantum-Readiness Working Group (QRWG) of the Canadian Forum for Digital Infrastructure Resilience (CFDIR), 2021). These protocols are indispensable for ensuring secure communication and are extensively employed across diverse domains, encompassing e-commerce, financial services, and government communications.

The risks posed by quantum computers to network protocols can be categorized into two primary areas. Firstly, quantum computers have the capability to compromise the encryption algorithms integral to these protocols, putting data confidentiality and integrity in jeopardy. In particular, encryption methods that rely on the complexity of mathematical problems, such as factoring large numbers, become vulnerable to quantum algorithms like Shor's algorithm (Shor, 1997). Secondly, quantum computers can undermine the authenticity and non-repudiation provided by digital signatures, which are essential for verifying the source and integrity of digital documents and transactions. The potential for quantum computers to break the cryptographic underpinnings of digital signatures presents the risk of unauthorized access and tampering of digital data. To address these vulnerabilities, researchers have been exploring post-quantum cryptographic solutions that are resistant to attacks from quantum computers. These solutions are based on mathematical problems that remain computationally challenging even for quantum algorithms. Recent advancements in post-quantum cryptography include lattice-based (Bos et al., 2018; Lyubashevsky et al., 2017; Ducas et al., 2018; Buchmann et al., 2019; Buchmann et al., 2020; Buchmann et al., 2021), code-based (McEliece, 1978; Chou et al., 2020; Aguilar-Melchor et al., 2018; Melchor et al., 2018; Aragon et al., 2017), hash-based (Bernstein et al., 2015), and isogeny-based (Hankerson et al., 2006; Washington, 2008) cryptographic schemes. Evaluating the suitability of these solutions for network protocols is essential to ensure the future security of communication systems.

* Corresponding author.
*E-mail addresses:* yaser.baseri@umontreal.ca (Y. Baseri), vikas.chouhan@unb.ca (V. Chouhan), ahafid@iro.umontreal.ca (A. Hafid).

The objective of this research is to comprehensively assess the security vulnerabilities of network protocols in the post-quantum era. We delve into the specific components and mechanisms of each protocol that quantum attacks can exploit. Additionally, we evaluate the resilience of these protocols against various quantum algorithms, taking into account factors such as key sizes, computational complexity, and performance implications. The outcomes of this research provide valuable insights into the potential risks faced by network protocols in the post-quantum era and inform the development of appropriate countermeasures. By understanding the vulnerabilities introduced by quantum computing and exploring the effectiveness of post-quantum cryptographic solutions, we can enhance the resilience and security of network protocols, ensuring they can withstand the imminent advances in quantum computing.

### 1.1. Motivation

The rapid advancements in quantum computing have brought about an imminent and unparalleled threat to the security of network protocols that rely on classical cryptographic algorithms. Quantum computers are evolving at a remarkable pace and possess the astonishing capability to rapidly and effectively break widely-used encryption methods. This poses a multifaceted risk not only to the confidentiality, integrity, and authenticity of network communications but also introduces tangible real-world consequences that cannot be underestimated.

In today's digitally interconnected world, secure communication serves as the linchpin for a myriad of critical operations. Network protocols function as the guardians of sensitive information across diverse domains, spanning e-commerce, financial transactions, government communications, and intra-organizational data exchange. However, any breach in the security of these protocols carries the potential for a chain reaction of devastating repercussions. Financial institutions, e-commerce platforms, and businesses of all sizes stand exposed to the looming specter of substantial financial losses. Quantum adversaries armed with unprecedented computational power can compromise the encryption mechanisms safeguarding financial transactions, leading to unauthorized access, fraudulent activities, and significant financial harm. Beyond the financial sector, there looms a threat to privacy. Personal and sensitive data exchanged via insecure network protocols could be laid bare, resulting in identity theft, unauthorized access to confidential information, and a profound erosion of individual privacy rights. Additionally, the security of governments and nations hangs in the balance, as quantum attacks could compromise national security, disrupt vital services, and endanger the safety of citizens. Equally perilous is the erosion of trust in digital communication, a linchpin for international diplomacy, commerce, and everyday interactions. The vulnerability of network protocols to quantum attacks has the potential to undermine confidence in the digital ecosystem, thereby hampering economic growth and international cooperation.

The urgency of this research is underscored by the stark reality that quantum computing advancements are no longer a distant theoretical prospect; they are rapidly becoming a tangible reality. In the face of this technological revolution, proactive measures are imperative to safeguard the very foundations of secure communication. By meticulously assessing the vulnerabilities of network protocols in the era of quantum computing, we can devise strategies, develop robust countermeasures, and ensure the ongoing security and reliability of network communications.

In summary, this research is driven by an imperative to shield network protocols from the tangible and far-reaching consequences of quantum attacks. By gaining a comprehensive understanding of the vulnerabilities introduced by quantum computing and exploring effective countermeasures, our goal is to fortify the very bedrock of secure communication in a quantum-powered world. The urgency of this endeavor cannot be overstated, as the security of our digital future hinges on the actions we take today.

### 1.2. Contributions of the work

Our research makes several significant contributions to the field of quantum computing and network protocols:

- **Assessing Vulnerabilities:** We conduct a comprehensive analysis to identify security vulnerabilities resulting from the advent of quantum computing in network protocols. This examination scrutinizes the specific components and mechanisms of each protocol, providing insights into the risks these protocols face in the post-quantum era. We evaluate the feasibility of quantum attacks, the computational resources required, and the potential consequences of successful breaches. This information contributes to a clearer understanding of the security landscape surrounding network protocols.
- **Exploring Post-Quantum Cryptographic Solutions:** We delve into the realm of post-quantum cryptographic solutions and their applicability to network protocols. Our investigation encompasses recent advancements in lattice-based, code-based, and multivariate polynomial-based schemes. By exploring these solutions, we aim to discover effective ways to mitigate the security risks posed by quantum computing.
- **Investigating Feasibility and Effectiveness:** In response to the vulnerabilities introduced by quantum computing, this study explores the applicability and effectiveness of post-quantum cryptographic solutions for network protocols. We examine various post-quantum cryptographic algorithms and protocols, assessing their resilience against quantum attacks and practicality for real-world deployment.
- **Informing Countermeasures and Recommendations:** The findings of this research contribute to the development of countermeasures and recommendations for enhancing the security of network protocols in the post-quantum era. By understanding vulnerabilities, risks, and potential solutions, we provide guidance to stakeholders, including protocol designers, implementers, and policymakers. This guidance ensures the secure and reliable operation of network communications.

- **Considering Practical Implications:** Our research takes into account the practical implications of transitioning to post-quantum cryptographic solutions for network protocols. Factors such as computational complexity, performance considerations, and interoperability are considered. This practical perspective offers valuable insights into the adoption of secure and efficient post-quantum cryptographic algorithms.

By achieving these research objectives, our study enhances our understanding of the vulnerabilities introduced by quantum computing in network protocols. Moreover, it provides guidance for the development of resilient and secure communication systems in the post-quantum era.

### 1.3. Organization

The structure of this paper is as follows: Section 2 provides an overview of prior research in this domain, encompassing foundational work and recent contributions. Section 3 explores the impact of quantum algorithms on computing security, highlighting vulnerabilities introduced by quantum computing and efforts in post-quantum cryptography standardization. In Section 4, we present the results of our vulnerability assessment for each network protocol, discussing potential risks and their implications for secure communication. This section also delves into the significance of post-quantum cryptographic solutions, considering factors such as security, performance, and interoperability. Section 5 explores Hybrid Approach Protocols, ensuring seamless business continuity in network security. Finally, Section 6 serves as the conclusion, summarizing key findings, and proposing future research directions to address the challenges posed by quantum computing in the realm of network protocols.

## 2. Related works

The emergence of quantum computing has sparked significant research into the vulnerabilities of network protocols within the security landscape. This section provides an overview of prior research in this domain, encompassing foundational work and recent contributions.

### 2.1. Quantum computing and cryptanalysis

Quantum computing's theoretical foundations and their implications for traditional cryptographic algorithms have been rigorously examined. Of note, Shor's algorithm's ability to efficiently factor large numbers, potentially compromising widely-used public-key encryption methods like RSA, has garnered significant attention (Shor, 1994). Additionally, Grover's algorithm, with its unstructured search capabilities that threaten symmetric key encryption, has been explored (Grover, 1996). Recent studies continue to deepen our understanding of quantum computing's vulnerabilities (Ji et al., 2022).

### 2.2. Post-quantum cryptography

The emerging area of post-quantum cryptography focuses on developing cryptographic solutions capable of withstanding quantum attacks. Recent notable contributions encompass lattice-based cryptography, code-based cryptography, hash-based, and isogeny-based schemes (Ding et al., 2017; Lyubashevsky, 2016). The NIST Post-Quantum Cryptography Standardization project has played a central role in advancing this research, providing a framework for evaluating and selecting post-quantum cryptographic algorithms (National Institute of Standards and Technology (NIST), 2023b). These endeavors pave the way for exploring alternative cryptographic methods capable of resisting the computational power of quantum computers (Li et al., 2023).

### 2.3. Vulnerability assessments of network protocols

Several studies have systematically assessed the vulnerabilities of network protocols in the context of quantum computing. These assessments focus on potential risks and attack scenarios posed by quantum adversaries. Recent research efforts have scrutinized the security of widely-used protocols, including SSH, TLS, and IPsec, pinpointing the vulnerabilities introduced by quantum attacks (Merli and Ursini, 2020; Katz and Schanzenbach, 2021; Zhang et al., 2020, 2023).

### 2.4. Transition strategies and hybrid approaches

In addition to identifying vulnerabilities, recent research has explored strategies for transitioning from classical to post-quantum cryptography. Hybrid approaches, which integrate classical and post-quantum cryptographic methods, have been proposed to ensure business continuity while mitigating risks during the migration process (Löhr and Meyer, 2020). Recent studies have assessed the feasibility and practicality of such transitions, considering factors such as performance, compatibility, and security (Wang et al., 2021).

### 2.5. Policy and standardization

The development of policies and standards for post-quantum cryptography has garnered significant attention, with a focus on recent contributions. Researchers have examined the role of organizations like NIST in shaping cryptographic standards, encompassing the criteria for evaluating post-quantum cryptographic algorithms and the standardization process itself (Lange, 2021). Recent studies contribute to a robust framework for adopting secure cryptographic practices in the post-quantum era (Mendel et al., 2022).

### 2.6. Real-world applications and case studies

An increasing body of research has extended beyond theoretical vulnerabilities to explore real-world applications and case studies. Recent investigations have delved into the practical implications of quantum-safe solutions in various domains, including finance, e-commerce, and government communications (Hou et al., 2021). These contemporary case studies offer valuable insights into the challenges and opportunities presented by quantum-safe network protocols in various operational contexts (Choi et al., 2023).

### 2.7. Quantum threat timelines and business continuity

Recent studies have provided unique perspectives on quantum threat timelines, with a focus on when quantum computers might pose cybersecurity risks (Mosca and Piani, 2022). Moreover, research reports have emphasized the importance of business continuity planning in the face of quantum computing's potential impact on cryptography systems (European Telecommunications Standards Institute (ETSI), 2016, 2020). Recommendations and guidelines for secure transitions between non-quantum-safe and quantum-safe states have also been explored in-depth (Mosca and Mulholland, 2017).

While the aforementioned related works provide valuable insights into the challenges posed by quantum computing in network security and offer recommendations for future research efforts, it is important to acknowledge that the landscape of quantum-safe network protocols is continually evolving. These studies, while comprehensive in their own right, may not cover every aspect of the rapidly changing field of quantum security. The growing importance of quantum-safe network protocols and the dynamic nature of this research domain underscores the need for a comprehensive survey to address the existing gaps. Recognizing these limitations, we have undertaken the task of writing this survey to provide a more exhaustive examination of vulnerabilities, risks, and migration strategies. Our aim is to offer a holistic view of quantum security challenges in networked environments, considering a diverse array of widely utilized security protocols and evaluating post-quantum cryptographic solutions. By doing so, we contribute to a more complete understanding of quantum computing's influence on network security.

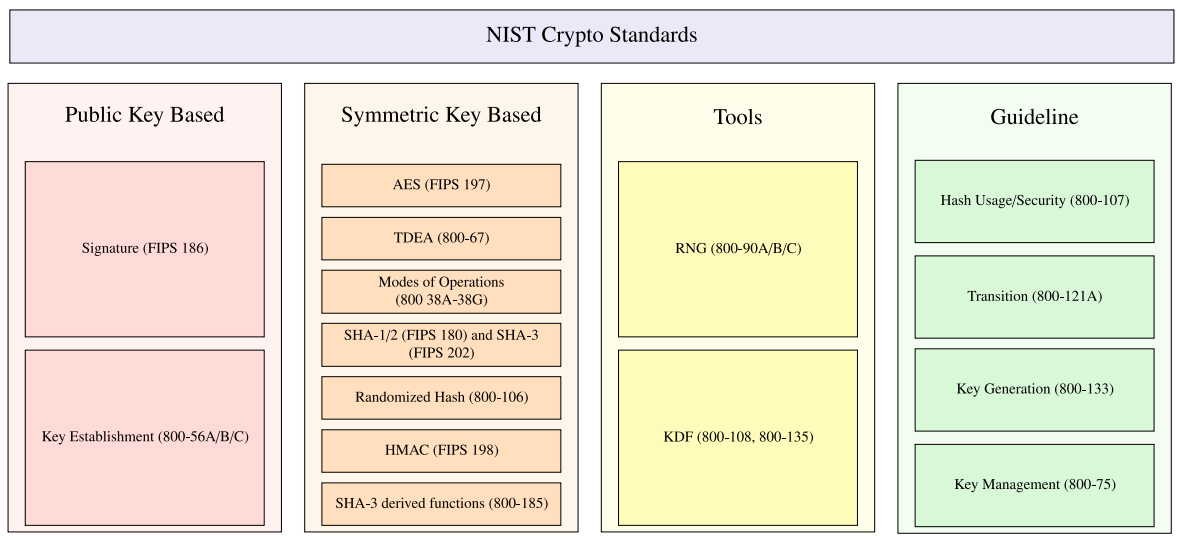

**Fig. 1.** NIST Crypto Standards.

In the following sections, we present a thorough investigation that not only identifies the challenges but also guides the development of practical countermeasures for securing networked environments in the quantum age. Our survey paper stands as a vital contribution to the ever-evolving field of quantum security, addressing the needs and concerns of protocol designers, implementers, and policymakers in the face of the advancement of quantum computing.

## 3. The transformative power of quantum algorithms in computing security

Quantum computing represents a groundbreaking frontier that harnesses the fascinating principles of quantum mechanics for computing tasks. In stark contrast to classical computers that rely on binary bits (0s and 1s), quantum computers utilize qubits. What makes qubits remarkable is their ability to exist in a superposition of both 0 and 1 states simultaneously. This distinctive property equips quantum computers with the potential to solve complex problems with unparalleled efficiency.

The true potential of quantum computing emerges through quantum algorithms, which can solve specific mathematical challenges exponentially faster than classical counterparts. An illustrative example of such a quantum algorithm is Shor's algorithm, which can factor large numbers exponentially faster than the best-known classical algorithms. This capability poses a potential threat to many cryptographic algorithms, including RSA, which forms the foundation of secure communication. These quantum algorithms manipulate and process information in fundamentally distinct ways, providing faster and more efficient solutions to intricate problems. This advancement promises to revolutionize fields like cryptography and optimization, offering computational capabilities that classical computers simply cannot match. Quantum computing's exponential speed advantage serves as the driving force behind its transformative impact on science, technology, and various industries.

### 3.1. Vulnerabilities introduced by quantum computing

The incredible computational power of quantum computers raises significant concerns about the security of network protocols relying on conventional cryptographic methods. Widely used protocols like TLS, IPsec, SSH, and PGP are secure against classical attacks but vulnerable to quantum threats. One of the most pressing vulnerabilities introduced by quantum computing is its potential to compromise established encryption techniques like RSA and ECC, which are pivotal for ensuring secure connections and data encryption. These techniques are integral components of standardization, such as those carried out by organizations like NIST (see Fig. 1). Quantum algorithms, notably Shor's algorithm, designed explicitly for quantum computers, have the capacity to undermine asymmetric encryption. This poses a substantial and imminent threat to secure communications and data protection.

Quantum computing also poses a grave threat to digital signatures, which rely on similar cryptographic principles as encryption. In particular, quantum computing, leveraging Shor's algorithm, threatens the authenticity and non-repudiation provided by digital signatures. Digital signatures serve to verify the source and integrity of messages, thus preventing unauthorized alterations. Shor's algorithm empowers attackers to retrieve private keys and forge digital signatures, elevating concerns regarding unauthorized access. Furthermore, quantum computing, through Grover's algorithm, threatens symmetric cryptography and hash functions. Grover's algorithm can search databases faster, potentially compromising data confidentiality and integrity. This has profound implications for data protection and secure communication.

In summary, quantum computing, driven by algorithms like Shor's and Grover's, poses substantial challenges to network protocol security and cryptographic techniques. These developments emphasize the urgent need for post-quantum cryptographic solutions to safeguard sensitive information in the era of quantum computing.

### 3.2. Post-quantum cryptography standardization

To shield our public key cryptographic algorithms from the looming threat of Quantum Computing (QC) and transition to a quantum-safe cryptographic state, we need quantum-resistant cryptographic algorithms. The National Institute of Standards and Technology (NIST) has taken the initiative to standardize such quantum-resistant cryptographic algorithms. This move is in response to the vulnerabilities that quantum computers pose to current cryptographic methods. The program includes a competition to identify post-quantum cryptographic algorithms. Post-quantum cryptography involves deploying quantum-safe cryptographic algorithms to secure Key Exchange Mechanisms (KEM)/Encryption (ENC) and signature algorithms against threats posed by QC.

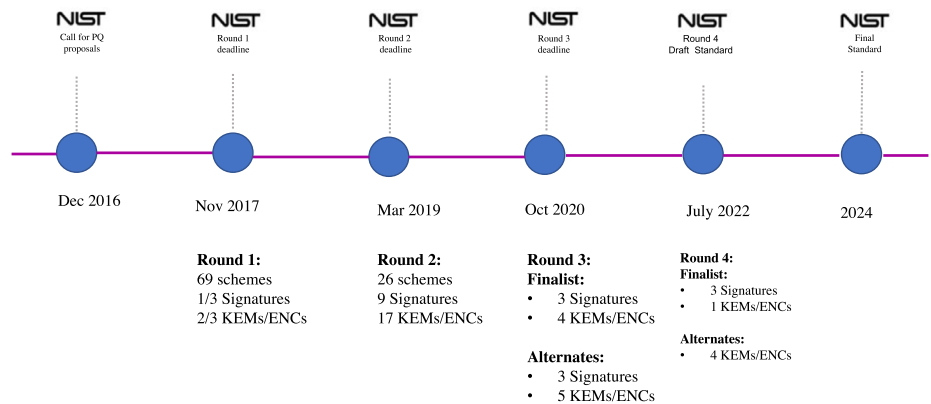

**Fig. 2.** NIST post-quantum cryptography timeline.

Several post-quantum cryptographic algorithms have been proposed, falling into categories like code-based, hash-based, lattice-based, and isogeny-based cryptographic algorithms. NIST is actively addressing the QC threat by soliciting post-quantum public-key exchange and digital signature algorithms. In 2022, NIST approved quantum-safe cryptographic candidates, including both KEM/ENC and Signatures for their fourth round of evaluations. The finalists, which include Kyber KEM (Bos et al., 2018), Dilithium (Lyubashevsky et al., 2017; Ducas et al., 2018), Falcon (Buchmann et al., 2019; Buchmann et al., 2020; Buchmann et al., 2021), and SPHINCS+ (Bernstein et al., 2015) signatures, were announced in July 2022. Four other algorithms have been chosen as alternates for the fourth round: McEliece (Chou et al., 2020), HQC (Aguilar-Melchor et al., 2018; Melchor et al., 2018), BIKE (Aragon et al., 2017), and SIKE (Hankerson et al., 2006; Washington, 2008). The NIST timeline for this initiative is depicted in Fig. 2.

### 3.3. Post-quantum cryptography: Vulnerabilities and weaknesses

Recent developments have exposed successful side-channel and cryptanalysis attacks on candidates in the NIST fourth round of evaluations, focusing on post-quantum cryptographic algorithms earmarked for quantum-safe cryptographic standards. This ongoing evaluation process continually unveils vulnerabilities, making it imperative to explore these attacks, potential safeguards, and the associated threats.

Quantum attackers, in their continuous effort to break post-quantum cryptographic algorithms, may employ various methods, primarily focusing on the recovery of secrets or the forging of signatures. They may exploit vulnerabilities like side-channels and mathematical analyses to breach the security of these algorithms. More specifically, side-channel attacks capitalize on information inadvertently disclosed during cryptographic algorithm execution, encompassing details like power consumption, electromagnetic radiation, or timing data. By dissecting this leaked information, attackers can illegally obtain sensitive data, including private keys. In contrast, cryptanalysis attacks center on undermining the encryption or signature schemes of a cryptographic algorithm. These attacks involve a comprehensive examination of the algorithm's structure and properties to pinpoint weaknesses that can be exploited for recovering confidential information. Table 1 provides an overview of the vulnerabilities that quantum attackers can exploit in their efforts to compromise the security of post-quantum cryptographic algorithms being considered for standardization in NIST's fourth round. Consequently, attackers have the means to execute both side-channel and mathematical analysis-based attacks, perpetuating the threat landscape.

## 4. Quantum-safe network and security protocols: Safeguarding against quantum threats

In this section, we conduct an in-depth security analysis of widely-used network and security protocols within the context of the four-layer TCP/IP model. These protocols are crucial for ensuring secure network communication and data protection, playing critical roles in various aspects of network security. They are commonly employed to safeguard data and communication across diverse domains, including e-commerce, financial services, and government communications (see Fig. 3).

Our primary focus is to evaluate the vulnerabilities introduced by the emergence of quantum computing within these protocols. We aim to identify specific security threats and potential attack vectors that may arise during the transition to quantum-resistant alternatives. To achieve this, we analyze the standardization and specifications of each protocol, examining their main components, intended purposes, recommended cryptographic suites, and the threats posed by quantum attackers. Subsequently, we explore quantum-resistant solutions to mitigate these threats and address any associated challenges and vulnerabilities that may surface during the implementation of these solutions. Our investigation begins with a comprehensive assessment of each protocol's vulnerabilities at different stages of protocol migration, encompassing both the currently widely used protocols and their quantum-resistant alternatives.

Security protocols are essential for organizations, providing critical measures for data source authentication, confidentiality, and data integrity. Established protocols in the TCP/IP model, such as SSH, TLS, and IPsec, which rely on public-key cryptographic algorithms, have been effective in defending against classical computer-based attacks. However, the advent of fault-tolerant quantum computers poses a significant threat to the security of these protocols, potentially compromising their cryptographic foundations within hours or even seconds, as detailed in a report by Deloitte (Scott Buchholz and Brown, 2021). Our investigation focuses on well-recognized standard security protocols documented in the Canadian National Quantum-Readiness guidelines (Quantum-Readiness Working Group (QRWG) of the Canadian Forum for Digital Infrastructure Resilience (CFDIR), 2021). We identify vulnerabilities that quantum adversaries might exploit during various phases of the security protocols. For each protocol and migration stage, we meticulously identify quantum computing-related threats using the STRIDE threat model. STRIDE is an acronym for Spoofing identity, Tampering with data, Repudiation threats, Information disclosure, Denial of service, and Elevation of privilege, which provides a systematic approach to threat modeling by identifying potential threats to system security (Kohnfelder et al., 2006; Shostack, 2014; Khalil et al., 2023). Subsequently, we present alternative quantum-resistant solutions for each protocol, highlighting their vulnerabilities and potential threats.

This section offers a comprehensive security analysis of network and security protocols in the context of quantum threats within the TCP/IP model. We assess the security threats and vulnerabilities introduced by quantum computing at various stages of these protocols. Our analysis explores attack vectors, potential risks, and alternative solutions. It also discusses the challenges and potential attacks associated with

**Table 1**

Attacks and countermeasures for NIST PQC 4th round candidates categorized by Cryptanalysis Attacks (CA), Timing Attacks (TA), Fault Attacks (FA), Simple Power Analysis (SPA), Advanced (correlation/differential) Power Analysis (APA), Electromagnetic Attacks (EM), Template Attacks (TMP) and Cold-Boot Attacks (CB).

| PQ Approaches | Finalists | | Alternates | | Available attacks and countermeasures* | | | | | | | | Possible countermeasures |
|---|---|---|---|---|---|---|---|---|---|---|---|---|---|
| | KEM/ENC | Signatures | KEM/ENC | Signatures | CA | TA | FA | SPA | APA | EM | TMP | CB | |
| Lattice-based | Kyber | | | | | | ✓✓ | ✓✓ | ✓✓ | ✓✓ | ✓ | ✓✓ | • **To defend against FA:** Mask the decryption process (Oder et al., 2016; Ravi et al., 2020a), or check the secret and error components of the LWE instances (Ravi et al., 2019b). • **To prevent SPA:** Mask the input (Hamburg et al., 2021), randomize the order of executed operations (Pessl and Primas, 2021). • **To avoid APA:** Mask the Number Theoretic Transform (NTT) (Pessl and Primas, 2019) (no countermeasures in place for the attacks mentioned in Dubrova et al. (2022)). • **To evade EM:** Mask the ECC procedures (Ravi et al., 2020b; Oder et al., 2016), FO transform operations (Ravi et al., 2020b), secure the secret key (Xu et al., 2021), discard ciphertexts (Xu et al., 2021), or randomly split secrets (Xu et al., 2021). • **For avoiding TMP,** no specific countermeasures for the attacks described in Ravi et al. (2021). • **To block CB:** Store the secret in the time domain instead of the frequency domain Albrecht et al. (2018). |
| | | Dilithium | | | | | ✓✓ | | ✓✓ | ✓✓ | | | • **To defend against FA:** (a) Check the secret and error components of the LWE instances (Ravi et al., 2019b), (b) Employ generic countermeasures such as double computation, verification-after-sign, and add additional randomness (Bruinderink and Pessl, 2018). • **To avoid APA:** Implement masking techniques, including linear secret sharing (Migliore et al., 2019) and use Boolean and arithmetic masking (Marzougui et al., 2022). • **To evade EM:** (a) Re-order operations and embed the vulnerable addition operation within the signing procedure (Ravi et al., 2019a), (b) Utilize it-slicing design for Number Theoretic Transform (NTT) (Singh et al., 2022). |
| | | Falcon | | | | ✓✓ | ✓✓ | ✓✓ | | ✓✓ | | | • **To avoid TA:** Enhance random shuffling through the Blind-Vector algorithm (McCarthy et al., 2019), or implement sample discarding (McCarthy et al., 2019). • **To defend against FA:** Duplicate computation of the signature (McCarthy et al., 2019), immediately verify after signing (McCarthy et al., 2019), and perform a zero-check on the sampled vector (McCarthy et al., 2019). • **To prevent SPA:** Effectively reduce the Hamming weight gap. • **To evade EM:** Conceal by maintaining a constant power consumption (Karabulut and Aysu, 2021), or mask by randomizing the intermediate values (Karabulut and Aysu, 2021). |
| Code-based | | | McEliece | | ✓✓ | ✓✓ | ✓✓ | ✓✓ | ✓✓ | ✓ | ✓✓ | ✓ | • **To prevent CA:** Increase the binary code length (Bernstein et al., 2008a) or use a decoding list to increase weight $w$ (Bernstein et al., 2008a). • **To defend against TA:** Artificially increase the degree of error locator polynomial for degrees lower than a threshold (Strenzke et al., 2008). • **To defend against FA:** Check decryption map output weight (Kreuzer and Danner, 2020), re-encrypt, and compare (Kreuzer and Danner, 2020) (no countermeasures in place for the attacks mentioned in Cayrel et al. (2020)). • **To counteract SPA:** Eliminate power trace patterns, branch statements, and data dependencies, maintain consistent power consumption and execution time (Petrvalsky et al., 2015) (no countermeasures in place for the attacks mentioned in Guo et al. (2022b)). • **To prevent APA:** Employ parallelization, shuffle (Jedlicka et al., 2022; Chen et al., 2015), or mask the cryptosystem by adding Goppa codewords to the ciphertext during permutation (Jedlicka et al., 2022; Petrvalsky et al., 2016). • **For avoiding EM,** no countermeasures in place for the attacks mentioned in Lahr et al. (2020). • **To block TMP:** Eliminate the memory access dependency on the content of the lookup-table and achieve constant runtime (Strenzke et al., 2008). • **For avoiding CB,** no countermeasures in place for the attacks mentioned in Polanco (2019). |
| | | | BIKE | | | ✓✓ | ✓✓ | | | | | | • **To prevent TA:** Increase the initial byte generation, eliminate additional random data generation calls, and implement constant-time random number generation techniques (Guo et al., 2022s). • **To defend against FA:** Employ countermeasures such as default failing, assembly-level instruction duplication, and introducing random delays in the system (Xagawa et al., 2021) (no countermeasures in place for the attacks mentioned in Cayrel et al. (2020)). |
| | | | HQC | | | ✓✓ | ✓✓ | ✓ | | ✓✓ | | | • **To avoid TA:** Increase the initial byte generation, eliminate additional random data generation calls, and implement constant-time random number generation techniques (Guo et al., 2022a). Also, construct a constant-time decoding algorithm (Wafo-Tapa et al., 2020) (no countermeasures are in place for the attacks mentioned in Guo et al. (2020)). • **To defend against FA:** Employ countermeasures such as default failing, assembly-level instruction duplication, and introducing random delays in the system (Xagawa et al., 2021) (no countermeasures in place for the attacks mentioned in Cayrel et al. (2020)). • **For avoiding TMP,** there are no countermeasures in place for the attacks mentioned in Schamberger et al. (2020). • **To evade EM:** Implement masking using linear secret sharing (Goy et al., 2022). |

*(continued on next page)*

**Table 1** (*continued*).

| PQ Approaches | Finalists | | Alternates | | Available attacks and countermeasures* | | | | | | | | Possible countermeasures |
|---|---|---|---|---|---|---|---|---|---|---|---|---|---|
| | KEM/ENC | Signatures | KEM/ENC | Signatures | CA | TA | FA | SPA | APA | EM | TMP | CB | |
| Hash-based | | | | SPHINCS+ | | | ✓✓ | | ✓✓ | | | | • **To defend against FA**: (a) Implement redundant signature computation for resilience (Castelnovi et al., 2018), (b) derive few-time signature (FTS) indices from public values, not secret ones (Castelnovi et al., 2018), (c) establish inter-layer links in the hyper-tree for fault detection in tree computation, preventing invalid signatures (Castelnovi et al., 2018), (d) employ fault detection through sub-tree recomputation with node swaps and an enhanced fault-resistant hash function (Genêt et al., 2018), (e) generate and store one-time signatures for efficient reuse (Genêt et al., 2018), (f) verify vulnerable instructions by recomputing on diverse hardware modules to identify discrepancies (Genêt et al., 2018).<br>• **To prevent APA**: Hide the order of the Mix procedures (Kannwischer et al., 2018). |
| Isogeny-based | | | SIKE | | ✓ | | ✓✓ | | ✓✓ | ✓✓ | | ✓ | • **For avoiding CA**, no countermeasures in place for the attacks mentioned in Castryck and Decru (2022).<br>• **To evade FA**: Employ countermeasures such as default failing, assembly-level instruction duplication, and introducing random delays for variable protection, as well as pushing curves through isogenies and enhancing the probability of recovering correct elliptic curve coefficients during key generation while verifying the implementation afterward (Xagawa et al., 2021; Tasso et al., 2021).<br>• **To counteract APA**: Use of a supersingular curve $E_A$ that generates points of order $3^{e_3}$ in $E_A[3^{e_3}]$ (De Feo et al., 2022).<br>• **To prevent EM**: Use of a supersingular curve $E_A$ that generates points of order $3^{e_3}$ in $E_A[3^{e_3}]$ (De Feo et al., 2022).<br>• **For avoiding CB**, no countermeasures in place for the attacks mentioned in Villanueva-Polanco and Angulo-Madrid (2020). |

* ✓: Attacks are feasible with no countermeasure in place, ✓✓: Attacks can be mitigated or prevented by effective countermeasures, ✓✗: Attacks are viable, but countermeasures are only partially available.

| TCP/IP Model Layers | Function | Security Protocols |
|---|---|---|
| Application Layer<br>Protocols: HTTP, FTP, SMTP, POP3 | Manages user interfaces, data exchange, and network services to applications. | SSH, SFTP, FTPS, DNSsec, SAML, OAuth, Kerberos, LDAP, PGP, S/MIME |
| Transport Layer<br>Protocols: TCP, UDP | Ensures reliability and provides error-checking mechanisms. | TLS, mTLS |
| Internet Layer<br>Protocols: IP, ICMP, BGP | Handles routing and forwarding of data packets across the network. | IKE, IPsec |
| Network Interface Layer<br>Protocols: Ethernet, PPP | Manages the physical transmission of data over the network medium. | WiFi/WPA, DECT |

**Fig. 3.** Network and security protocols aligned with TCP/IP model layers.

these solutions and examines emerging threats in the quantum era. To facilitate our analysis, we provide detailed descriptions of each layer of the TCP/IP model and its corresponding security protocols:

- **Application Layer:** This layer provides an interface for network services and applications. It includes protocols such as HTTP, FTP, SMTP, and POP3, each serving different communication purposes like web browsing, file transfer, and email services. Security protocols here include SSH for secure remote logins, SFTP and FTPS for secure file transfers, DNSsec for domain name security, SAML and OAuth for authentication and authorization, and Kerberos for network authentication.
- **Transport Layer:** This layer is responsible for providing communication services directly to the application processes running on different hosts. It uses TCP for reliable data transmission and UDP for quicker, connectionless communication. To secure data transfer, it employs TLS for encryption and mTLS for mutual authentication.
- **Internet Layer:** This layer is responsible for routing data, addressing, and WAN delivery. Protocols include IP for packet forwarding, ICMP for diagnostics, OSPF for routing within a domain, and BGP for routing between domains. Security is provided by IKE for VPN connections and IPsec for data security, focusing on integrity, confidentiality, and authentication.
- **Network Interface Layer:** The final layer handles physical addressing, LAN delivery, and bit transmission. Protocols such as Ethernet and PPP are used here. Security protocols include those for securing WiFi/WPA for wireless security and DECT for secure cordless communications, ensuring the protection of data over physical connections.

In the subsequent sections, we conduct an in-depth examination of various network and security protocols within the TCP/IP model. This analysis focuses on their vulnerability to quantum computing threats, exposes potential weaknesses, and proposes remedies for transitioning to quantum-safe cryptographic methods. We encompass aspects such as standardization, essential components, primary objectives, recommended cipher suites, and the existing quantum-related risks these protocols face. Furthermore, we explore alternative quantum-resistant solutions for these protocols, providing insights into the associated challenges and vulnerabilities during their implementation. We also explain the possible threats that these quantum-resistant alternatives might encounter and present corresponding countermeasures for their mitigation. Subsequent sections will delve into the specific purposes of each protocol, and delve into the available or potential quantum-resistant solutions for each one.

### 4.1. Application layer

The Application Layer, situated within the TCP/IP model, holds a pivotal role as the user interface and gateway for applications in network communication. It serves as the vital bridge between users and the network, managing user interfaces and offering essential network services to applications. This layer plays a key role in ensuring the seamless exchange of data among diverse software applications, including web browsers, email clients, and file transfer programs. Its core responsibility lies in presenting data to applications in a format they can comprehend while efficiently supervising communication between various software running on different devices. Moreover, the Application Layer defines the protocols and services that applications use to communicate with the lower layers of the network stack, establishing itself as the cornerstone of network communication.

At the heart of this layer lie essential protocols such as HTTP, FTP, SMTP, and POP3, enabling communication with a myriad of network services. Additionally, a range of security protocols, including SSH, SFTP, FTPS, DNSsec, SAML, OAuth, and Kerberos, play a fundamental role in safeguarding data. However, the emergence of quantum computing technology has raised concerns regarding the security of these protocols. To address this challenge and ensure the ongoing confidentiality and integrity of network communications, it is imperative to explore quantum-resistant alternatives.

In the following sections, we will delve into the network and security protocols of the Application Layer and examine potential quantum-resistant solutions. As quantum computing capabilities evolve, the security of these protocols becomes increasingly significant due to the potential vulnerabilities posed by quantum attacks. Thus, exploring quantum-resistant cryptographic methods tailored to withstand the capabilities of quantum adversaries is essential to ensure the continued security and integrity of data presentation and transmission in the Application Layer.

#### 4.1.1. SSH protocol

SSH (Secure Shell) is a pivotal protocol in the realm of network security, serving a multifaceted role. It comprises three foundational sub-protocols: the Transport Layer protocol, user authentication protocol, and connection protocol, as meticulously detailed by Schneier (2020). Its significance lies in its ability to fulfill the crucial need for secure remote login and the protected transmission of network services across potentially compromised environments.

SSH achieves its security objectives through a fusion of cryptographic techniques, leveraging both Public-Key Cryptography (such as EdDSA, ECDSA, RSA, and DSA) and Symmetric Cryptography (including AES, RC4, and ChaCha20-Poly1305), all meticulously outlined in the SSH RFC (Lonvick and Ylonen, 2021). These cryptographic mechanisms are instrumental in guaranteeing data authenticity, integrity, and confidentiality during transmission. However, it is imperative to recognize that SSH's security is not immune to the looming threat of quantum computing. Public-key cryptography algorithms employed by SSH, such as EdDSA, ECDSA, RSA, and DSA, are susceptible to compromise through Shor's algorithm (Shor, 2022). Simultaneously, symmetric ciphers like AES and SHA1/SHA2 may face potential vulnerabilities from Grover's algorithm (Grover, 2023). These quantum-related concerns introduce the specter of threats such as spoofing, tampering, and information disclosure, posing substantial risks to the security of data traversing networks secured by SSH.

To protect against quantum attacks, SSH can adopt quantum-resistant solutions like OpenSSH (Author's Name, 2022; Quantum-Safe OpenSSH Team, 2023) and OQS-libssh (Open Quantum Safe (OQS), 2022, 2024). However, these implementations come with certain challenges and vulnerabilities. One challenge is the potential for significant communication overhead when using quantum-resistant algorithms.

This overhead may lead to network congestion and fragmentation issues, triggering retransmissions. Additionally, the use of quantum-resistant protocols opens up the possibility of denial-of-service (DoS) attacks and data exfiltration. Furthermore, side-channel and mathematical analysis attacks, as mentioned in Table 1, could lead to information disclosure.

To mitigate these issues, several countermeasures can be implemented. Congestion control mechanisms should be employed to manage communication overhead associated with quantum-resistant algorithms and prevent network congestion (Zhou et al., 2022; Nicula and Zota, 2019). This helps regulate data flow, preventing network congestion. To address fragmentation and retransmission issues, careful consideration of message size is essential when using quantum-resistant protocols (IBM, 2023). Fragmenting messages into smaller, manageable segments and ensuring proper reassembly at the destination can prevent fragmentation and retransmission issues. Implementing verification mechanisms for fragment integrity and origin can enhance security by confirming the authenticity and completeness of message fragments. To mitigate communication overhead and prevent DoS attacks, providing ample bandwidth to SSH connections is crucial. This ensures the network can handle increased traffic. Building redundancy into the network infrastructure provides failover options in case of congestion issues. Deploying hardware and software modules to counter Distributed Denial-of-Service (DDoS) attacks protects against attempts to flood the network with traffic, ensuring SSH service availability (Garcia and Blandon, 2022a; Liu et al., 2018). To counter side-channel and mathematical analysis attacks, the solutions outlined in Table 1 can be employed (Peikert, 2017; Bernstein et al., 2017). These measures collectively aim to safeguard SSH connections in a post-quantum computing landscape, protecting against threats like information disclosure and DoS attacks.

### 4.1.2. SFTP protocol

Secure File Transfer Protocol (SFTP) is a crucial protocol for securely accessing, transferring, and managing files over unreliable networks, utilizing the Secure Shell (SSH) framework (Friedl and Ylonen, 2021; Barrett et al., 2008). Its primary objectives are to provide client/server authentication, data integrity, and confidentiality. SFTP achieves these goals by employing various cryptographic suites. However, it is important to note that the protocol is vulnerable to quantum computing threats (Smith, 2022).

Regarding the recommended cipher suites, SFTP relies on Public-Key Cryptography mechanisms such as RSA, DSS, and DH, all of which are vulnerable to quantum computer attacks, notably Shor's algorithm (Shor, 1997). Additionally, SFTP employs symmetric ciphers like 3DES, Blowfish, Twofish, and others, all of which are weakened in a quantum computing environment by Grover's algorithm (Grover, 2023). Consequently, quantum computing introduces novel threats to SFTP, encompassing spoofing, tampering, and information disclosure. The potential for quantum computers to compromise the encryption schemes used in SFTP may jeopardize the confidentiality and integrity of data transmitted via this protocol. Therefore, it is imperative to explore quantum-resistant cryptographic solutions or post-quantum cryptography to mitigate these vulnerabilities and ensure the ongoing security of SFTP in the quantum era (National Institute of Standards and Technology (NIST), 2023a).

Quantum-resistant SFTP is a critical development in the realm of secure data exchange. To address the looming threat of quantum computing, quantum-resistant SFTP can be evolved with innovative solutions. One approach is to replace the traditional SSH protocol with quantum-resistant alternatives like OpenSSH (OpenSSH Development Team, 2024) and OQS-libssh (oqs, 2022). While these solutions are robust against quantum attacks, they do introduce some challenges. These challenges include significant communication overhead, potential network congestion, fragmentation issues leading to retransmissions, susceptibility to denial-of-service attacks, and the risk of data exfiltration and information disclosure through side-channel and mathematical analysis attacks.

To mitigate these challenges, several countermeasures can be employed. Congestion control mechanisms can be implemented to manage network congestion effectively. Additionally, careful consideration of message size, especially in scenarios where quantum-resistant protocols are adopted, can help prevent fragmentation issues and retransmissions. To handle the increased communication overhead and protect against denial-of-service attacks, organizations can invest in enhanced network infrastructure, redundancy, and DDoS resilience hardware and software modules. Furthermore, specific solutions mentioned in Table 1 can be employed to safeguard against side-channel and mathematical analysis attacks.

However, it is essential to remain vigilant as quantum threats like information disclosure and denial of service remain persistent concerns in the quantum-resistant SFTP landscape. Therefore, organizations should continuously monitor and adapt their security measures to stay ahead of potential vulnerabilities in this ever-evolving environment (Dempsey et al., 2011).

### 4.1.3. FTPS protocol

FTPS, which stands for File Transfer Protocol Secure, is designed to provide security support for the classic FTP protocol by incorporating Transport Layer Security (TLS). This enhancement aims to ensure the confidentiality, integrity, and authentication of data during file transfers (Ford-Hutchinson and Adams, 2005). FTPS employs a combination of public-key cryptography, such as RSA and ECDH, and symmetric cryptography, like AES and SHA2, to protect data in transit. However, it is important to note that classic FTPS, like many other cryptographic protocols, is vulnerable to quantum computing threats. Quantum computers, leveraging algorithms like Shor's and Grover's, pose a significant risk to the cryptographic mechanisms employed by FTPS. Shor's algorithm can break the public-key cryptography used in FTPS, potentially compromising authentication and confidentiality. On the other hand, Grover's algorithm can weaken the symmetric encryption schemes, making data less secure. In summary, FTPS serves as a secure means of transferring files over a network, employing a combination of cryptographic techniques. Nevertheless, it faces vulnerabilities when confronted with the capabilities of quantum attackers, highlighting the need for ongoing research and development in post-quantum cryptography to ensure data security in the quantum era.

Quantum-resistant FTPS is a crucial development in the realm of secure data exchange. To address the imminent threat posed by quantum computers to cryptographic protocols, FTPS can be fortified with quantum-resistant solutions. For instance, FTPS can adopt the OQS-OpenSSL (Open Quantum Safe (OQS), 0000) or KEMTLS (Schwabe, 2023), replacing traditional TLS, to ensure quantum resilience. However, this transition is not without its challenges and potential vulnerabilities. FTPS may face issues such as significant communication overhead, network congestion, and fragmentation, which can trigger retransmission and potentially lead to denial of service attacks. Additionally, quantum-resistant FTPS must be prepared to tackle the threats of information disclosure and denial of service.

To mitigate these challenges and fortify the quantum resistance of FTPS, several countermeasures can be employed. Congestion control mechanisms can be implemented to prevent or alleviate network congestion. Furthermore, careful consideration of message size can help avoid fragmentation and retransmission issues. To combat the risk of information disclosure and denial of service, strategies such as providing ample bandwidth, redundancy in infrastructure, and deploying DDoS resilience hardware and software modules like firewalls can be adopted. Additionally, side-channel attacks and mathematical analysis threats can be addressed using solutions mentioned in the provided table.

In summary, quantum-resistant FTPS, achieved through the adoption of quantum-resistant protocols like OQS-OpenSSL or KEMTLS, is essential to safeguard data transfers in a post-quantum world. While challenges and threats exist, implementing countermeasures and best practices can ensure the security and reliability of quantum-resistant FTPS. This transition is a critical step in preparing for the era of quantum computing and maintaining the confidentiality and integrity of data transfers.

### 4.1.4. DNSSEC protocol

DNSSEC, or Domain Name System Security Extensions, plays a crucial role in enhancing the security of the Internet's domain name system. It is an essential protocol for securing the Domain Name System (DNS) by providing digital signatures for DNS data to ensure its authenticity and integrity. It achieves this by adding cryptographic signatures to existing DNS records, effectively protecting the DNS infrastructure from vulnerabilities and attacks (Arends et al., 2005a,c,b). DNSSEC primarily focuses on two key aspects: data origin authentication and data integrity protection.

To provide data origin authentication, DNSSEC employs various public-key cryptographic algorithms such as RSA, DSA, and ECDSA (Eastlake and Panitz, 2001). However, it is important to acknowledge that certain cryptographic algorithms are susceptible to quantum attacks, most notably Shor's algorithm, which poses a substantial threat to their security. This vulnerability opens the door to quantum attackers compromising the authenticity and integrity of DNS data, potentially leading to vulnerabilities in spoofing and tampering. Spoofing attacks seek to deceive users by presenting falsified DNS information, which can result in malicious destinations. Tampering involves unauthorized alterations of DNS records, leading to data corruption or redirection to malicious websites. Despite these vulnerabilities, DNSSEC remains an essential security protocol, offering a higher level of assurance regarding the authenticity and integrity of DNS data. However, as the threat of quantum computing grows, it is imperative to continually evaluate and adapt DNSSEC to address evolving quantum threats and classic security challenges to maintain the security and reliability of the Internet's domain name system (ICANN, 2020).

To address the quantum threat in DNSSEC, it is crucial to replace classic cryptographic signatures with post-quantum signature schemes (Bernstein et al., 2008b). PQ signatures are designed to withstand attacks from quantum computers, which have the potential to break traditional cryptographic algorithms. These PQ signatures should be paired with longer symmetric keys and hash functions that are quantum-resistant, ensuring the security of DNSSEC data (National Security Agency (NSA), 2016). One of the challenges in implementing quantum-resistant DNSSEC is the limited packet size in DNSSEC messages, which is typically 512 bytes. DNSSEC additional information like signatures and keys must fit within this constraint. When transitioning to post-quantum (PQ) cryptography, the larger size of PQ signatures and keys must be accommodated without causing fragmentation issues. This can be achieved by optimizing PQ algorithms for smaller signature sizes or by modifying DNSSEC message sizes to accommodate PQ signatures. For the symmetric key cryptography that DNSSEC uses, longer keys should be used (Müller et al., 2020; Beernink, 2022). Another concern is the risk of denial-of-service (DoS) attacks targeting DNSSEC. To mitigate these attacks, DNS infrastructure should be designed to handle the increased computational load associated with PQ cryptography. This may involve deploying additional computational resources, such as hardware acceleration or distributed processing, to maintain DNSSEC's performance and responsiveness in the face of attacks (Garcia and Blandon, 2022a; Liu et al., 2018). Furthermore, it is essential to address the potential for info disclosure in DNSSEC due to side-channel attacks. Implementing countermeasures against side-channel attacks, such as cache side-channel protection and timing attack mitigation, is critical to maintaining the confidentiality of DNSSEC operations (see Table 1).

In summary, quantum-resistant DNSSEC involves replacing classic signatures with PQ signatures, accommodating larger signature sizes within DNSSEC message limits, and addressing potential DoS and info disclosure risks through infrastructure enhancements and side-channel attack protection. This transition will help ensure the continued security of DNS operations in the post-quantum era.

### 4.1.5. SAML protocol

Classic Security Assertion Markup Language (SAML), denoted as SAML (v2), plays a crucial role in providing authentication to multiple applications and supporting both authentication and authorization (Campbell and Mortimore, 2015). It relies on cryptographic techniques, utilizing public-key cryptography such as RSA and DSA, although these algorithms are susceptible to attacks by quantum computers, like Shor's algorithm. Additionally, SAML employs symmetric cryptography, particularly SHA2, which could be weakened by Grover's algorithm. While SAML serves its purpose in authentication and authorization, its vulnerability to quantum attacks, particularly spoofing and elevation of privilege, raises concerns in an era where quantum computing capabilities are advancing. Therefore, it is imperative for organizations relying on SAML to consider quantum-resistant alternatives and quantum-safe encryption methods to mitigate these potential threats effectively.

Quantum-resistant SAML introduces a new era of secure authentication and authorization in the post-quantum world. In this paradigm shift, traditional public-key algorithms like RSA or DSA, which were the cornerstones of SAML, should be replaced by robust post-quantum cryptographic solutions. These new PQ algorithms enable SAML to resist quantum attacks that could compromise the security of classical encryption. However, this transition comes with its own set of challenges and potential vulnerabilities. The adoption of PQ algorithms can introduce significant communication overhead, leading to network congestion and fragmentation issues that may trigger retransmission. Moreover, the increased size of PQ public keys can impact system performance and create the potential for DoS attacks. Additionally, the threat of information disclosure looms large, with side-channel attacks exploiting observable timing differences and cache access patterns. To address these challenges, the quantum-resistant SAML protocol incorporates various countermeasures. Congestion control mechanisms are implemented to manage network traffic effectively (Jay et al., 2018; Bohloulzadeh and Rajaei, 2020; Jay et al., 2019). To mitigate fragmentation and retransmission issues, numerous strategies have been suggested in the literature. These strategies encompass optimizing message size, selecting suitable transport protocols like HTTPS, applying compression methods, configuring network infrastructure effectively, conducting routine testing and monitoring, and exploring alternative communication technologies (Summit360, 2019a,b). To prevent DoS attacks, it is essential to provide the necessary bandwidth, redundancy, and deploy DDoS resilience hardware and software. Protection measures such as firewalls and DDoS protection appliances are also essential for safeguarding the infrastructure (Garcia and Blandon, 2022a; Liu et al., 2018). To mitigate the risk of side-channel attacks, PQ algorithms with enhanced security features are adopted (see Table 1). In addition to these measures, the SAML protocol leverages quantum-resistant TLS solutions to protect against information disclosure and DoS attacks. The replacement of traditional certificates with PQ certificates further enhances the protocol's security posture.

Overall, quantum-resistant SAML is a pivotal development in the world of secure authentication and authorization. While it faces its unique challenges, the adoption of PQ algorithms and a comprehensive set of countermeasures ensures that SAML remains robust and resilient against quantum threats, safeguarding critical digital transactions and communications in the post-quantum era.

### 4.1.6. Oauth protocol

The OAuth Protocol, abbreviated as Open Authorization, is a widely adopted framework that enables websites and applications to access resources from other web services on behalf of a user without exposing their credentials (Hardt, 2012; Jones and Hardt, 2012; Lodderstedt et al., 2013; Richer et al., 2015; Parecki and Waite, 2015). It accomplishes this through various sub-protocols like Token Binding, Key Generation, and Proof of Possession. While OAuth primarily focuses on authorization, which is its core functionality, it also ensures token protection through signature mechanisms and may employ TLS for secure token transmission.

It is important to note that traditional OAuth implementations rely on cryptographic primitives that are susceptible to quantum attacks. For instance, RSA and ECDHE, utilized in OAuth, can be compromised by Shor's algorithm (Shor, 1997), and symmetric encryption algorithms like AES and SHA2, as well as ChaCha20-Poly1305 through TLS, are affected by Grover's algorithm (Grover, 2023). As quantum computing advances, these vulnerabilities pose a security risk to OAuth, potentially enabling attackers to gain unauthorized access and compromise sensitive data.

To address the quantum threat, the transition to a quantum-resistant OAuth involves significant changes. Traditional HMAC-SHA1 token signatures should be replaced with quantum-resistant alternatives, often requiring longer keys to enhance security against quantum attacks. Notably, key lengths may need to be significantly longer than 2048 bytes to ensure quantum resistance, given the evolving nature of cryptographic schemes in response to quantum advances. Additionally, the adoption of quantum-resistant OAuth necessitates a shift from traditional TLS to post-quantum TLS solutions to ensure the ongoing security of cryptographic mechanisms in a post-quantum world.

However, this transition to quantum-resistant OAuth is not without challenges. Like other quantum-resistant protocols, OAuth faces issues related to communication overhead, network congestion, and potential DoS attacks, mainly due to the use of quantum-resistant TLS. Larger keys for quantum resistance can increase communication overhead and lead to network congestion. To address these challenges, implementing congestion control mechanisms and ensuring adequate bandwidth for larger keys are crucial. Redundancy in infrastructure and the deployment of DoS resilience hardware and software modules, such as firewalls, can help mitigate DoS attacks (Hardt, 2018). To safeguard against potential side-channel attacks, OAuth implementations should incorporate solutions mentioned in Table 1.

Despite these security enhancements, it is important to acknowledge that quantum computing still presents risks, particularly in terms of information disclosure and DoS attacks. Organizations must remain vigilant in monitoring and updating their security measures to stay ahead of evolving threats in the quantum era (Hardt, 2018).

### 4.1.7. Kerberos protocol

Kerberos, a classic authentication protocol, plays a crucial role in securing access to systems over untrusted networks like the Internet. It achieves this by providing a mechanism for authenticating both clients and servers in client/server applications. Kerberos relies on symmetric cryptography, using algorithms like HMAC/AES, MD4, MD5, HMAC-SHA1/SHA2/SHA3, and CMAC/camellia for security (Bradner, 2005,?; Raeburn et al., 2005; Hartman et al., 2013; Zhu et al., 2011; Gutmann, 2017). However, it is important to note that these (only) *symmetric* cryptographic algorithms are vulnerable to attacks by quantum computers, such as Grover's algorithm. This vulnerability makes Kerberos susceptible to spoofing attacks, where an attacker could impersonate a legitimate user or server. To enhance the security of Kerberos against quantum threats, it is essential to consider post-quantum cryptographic algorithms that can withstand the power of quantum computers. Upgrading the encryption methods used within Kerberos is a critical step to maintain its effectiveness in an evolving threat landscape.

Quantum-Resistant Kerberos (PQ Kerberos) addresses the need for secure authentication and authorization in a post-quantum computing era. In PQ Kerberos, the focus is on mitigating quantum computing threats, particularly those posed by Grover's algorithm and the potential for quantum computers to break symmetric keys derived from passwords. To achieve quantum resistance, PQ Kerberos employs longer and strongly random symmetric keys for authentication instead of relying on password-derived keys, which may be susceptible to Grover's algorithm. This approach ensures that the symmetric keys used for authentication are robust against quantum attacks. By implementing these measures, PQ Kerberos aims to prevent spoofing attacks, where an attacker could impersonate a legitimate user. This quantum-resistant variant of Kerberos enhances the security of authentication protocols in a quantum-threat landscape. However, it is essential to balance security with practicality, as longer keys can also increase computational overhead and key management complexities, necessitating careful consideration during implementation.

### 4.1.8. LDAP protocol

LDAP (Lightweight Directory Access Protocol) is a crucial protocol used for providing directory services access and authorization, as well as maintenance for distributed directory information services over an IP network. Its primary purposes include facilitating secure access to directory information and supporting optional authentication via methods like binding with basic authentication or Simple Authentication and Security Layer (SASL) (Melrose and Dawson, 2006). LDAP also offers the option to enhance communication confidentiality and data integrity through the use of TLS (Melrose and Dawson, 2006). However, LDAP is not immune to quantum computing threats. Quantum computers, with their immense processing power, pose significant vulnerabilities to the security mechanisms used in LDAP. For instance, the public-key cryptography methods, including RSA and ECDH, which LDAP may employ via SASL, can be compromised by Shor's algorithm when executed on a sufficiently powerful quantum computer. Similarly, the symmetric cryptography techniques such as AES, SHA2, and ChaCha20-Poly1305, often used with TLS, are weakened by Grover's algorithm. These quantum threats put LDAP at risk of various security issues, including spoofing (via credential or SASL (Melnikov, 2006)), tampering (when TLS is optionally used), information disclosure (also related to TLS), and potentially an elevation of privilege. To maintain the security of LDAP in the face of quantum computing advances, it becomes imperative to explore post-quantum cryptography solutions and employ robust security measures.

Quantum-resistant LDAP presents a robust solution to secure directory services against potential quantum threats. In this context, several aspects must be addressed. To counter the threat of quantum computing, LDAP can adopt longer and more secure symmetric keys or passwords for SASL authentication. These keys should possess sufficient entropy to withstand birthday attacks and Grover's algorithm, ensuring that the authentication process remains robust. Additionally, the use of TLS can be enhanced with quantum-resistant TLS solutions like OQS-OpenSSL or KEMTLS to protect data in transit from quantum attacks. PQ certificates should also replace classic certificates to fortify the Public Key Infrastructure (PKI).

However, several challenges and potential attacks must be mitigated. The optional use of a password or symmetric key in SASL can be vulnerable to birthday attacks and Grover's algorithm, making it necessary to ensure that these secrets are generated securely. Moreover, the use of TLS, while offering encryption, can be susceptible to attacks similar to those mentioned for quantum-resistant TLS, such as network congestion and denial of services. To address these concerns, congestion control mechanisms should be implemented to manage network traffic effectively. Furthermore, protecting against potential attacks, such as spoofing, info disclosure, and denial of services, is crucial. Robust countermeasures like securing long-term symmetric keys or passwords for each user and employing solutions from the quantum-resistant

TLS context can enhance security. Additionally, measures to mitigate side-channel attacks, as detailed in Table 1, should be adopted.

In summary, quantum-resistant LDAP offers a fortified LDAP system that uses longer and more secure symmetric keys or passwords, replaces TLS with quantum-resistant TLS solutions, and strengthens PKI with PQ certificates. These measures help mitigate potential quantum threats and ensure the integrity, confidentiality, and availability of directory services.

### 4.1.9. PGP protocols

PGP, or Pretty Good Privacy, is a well-established encryption protocol renowned for its pivotal role in safeguarding privacy and authentication in data communication, particularly within encrypted emails and files (Callas et al., 2007b; Schiller, 1996). It achieves these objectives through a sophisticated blend of cryptographic techniques, incorporating public-key cryptography methods such as RSA, Elgamal, DH, and DSA. However, it is essential to recognize that these public-key cryptography algorithms are susceptible to potential vulnerabilities posed by quantum computers, specifically Shor's algorithm, which could potentially compromise the security of PGP-encrypted data.

In addition to its reliance on public-key cryptography, PGP also employs symmetric cryptography algorithms like MD5, SHA1, CAST, IDEA, or Triple-DES. While these symmetric algorithms have demonstrated effectiveness in the classical computing realm, they do face vulnerabilities from quantum computers, notably Grover's algorithm, which can substantially weaken their security.

Despite its reputation for data privacy and authentication, PGP is not impervious to quantum-related vulnerabilities. Quantum attackers could potentially exploit these vulnerabilities to engage in spoofing, data tampering, repudiation of actions, and even unauthorized information disclosure. As quantum computing capabilities continue to advance, it becomes imperative to explore and implement post-quantum cryptographic solutions to fortify data transmitted through PGP and similar protocols against emerging quantum threats.

The transition towards quantum-resistant PGP represents a pivotal development in the realm of secure communication. Traditional PGP, originally reliant on classic public-key cryptography, should adapt to post-quantum solutions to counter the looming threat of quantum computers. However, this transition is not without its set of challenges. Migrating from PGP's traditional public keys, which are limited to a maximum size of 4096 bits, to post-quantum cryptography presents several difficulties. These challenges encompass significant communication overhead, potential network congestion, fragmentation issues leading to retransmissions, and the risk of DoS attacks. Additionally, adversaries might exploit side-channel attacks, such as cache side-channel attacks, to gain access to private keys. Addressing these issues necessitates the implementation of various countermeasures.

To effectively manage network resources and prevent congestion, the integration of congestion control mechanisms is essential. Furthermore, careful consideration must be given to message size during the transition to post-quantum cryptography to prevent fragmentation and the subsequent retransmission of data. Ensuring ample bandwidth, fortifying infrastructure redundancy, and deploying DDoS resilience measures such as firewalls and protection appliances are vital steps in thwarting potential DoS attacks.

Bolstering defenses against side-channel attacks and cache side-channel vulnerabilities is equally crucial. Adopting security measures from the realm of post-quantum cryptography is imperative, as these measures are purposefully designed to safeguard sensitive information and uphold confidentiality.

The combination of quantum-resistant PGP and the countermeasures presented here fortify the solution against potential challenges and vulnerabilities. This comprehensive approach helps mitigate quantum threats, ensuring protection against spoofing, tampering, repudiation, information disclosure, and DoS attacks. Quantum-resistant PGP represents a significant stride towards ensuring the continued privacy and security of digital communications in an era marked by the rapid advancement of quantum computing capabilities.

### 4.1.10. S/MIME protocol

S/MIME, an acronym for Secure/Multipurpose Internet Mail Extensions, represents a protocol engineered to fortify electronic messaging with privacy and data security. It encompasses various critical objectives, including authentication, integrity, non-repudiation, and confidentiality, all of which are achieved through cryptographic techniques (Internet Engineering Task Force, 2018). However, it is vital to acknowledge that classic S/MIME implementations are susceptible to quantum attackers.

In the realm of classic S/MIME, public-key cryptography methods such as RSA and DSA are employed for encryption and digital signatures. Regrettably, these cryptographic techniques are vulnerable to quantum computers wielding Shor's algorithm. Furthermore, symmetric cryptography, notably AES, which secures the actual message content, is compromised by Grover's algorithm when executed on a quantum computer. Consequently, classic S/MIME confronts a spectrum of quantum threats, encompassing spoofing, tampering, repudiation, and information disclosure. To bolster the security of S/MIME in the post-quantum landscape, it becomes imperative to explore and implement quantum-resistant encryption techniques and cryptographic algorithms. This proactive approach ensures that S/MIME retains its potency in safeguarding electronic communications in the face of quantum computing advancements (Internet Engineering Task Force, 2018; Döberl et al., 2023).

Quantum-resistant S/MIME emerges as a pivotal advancement in fortifying email communication against the perils posed by quantum adversaries. In this context, S/MIME protocols can be revamped to withstand potential quantum risks by substituting classic public-key cryptography with post-quantum alternatives. The incorporation of PQ public-key cryptographic algorithms and the utilization of extended symmetric keys constitute the foundations of a potential quantum-resistant S/MIME solution, reinforcing the security of email communications in the presence of quantum threats. A distinguishing characteristic of quantum-resistant S/MIME is the elimination of constraints on public key size, enabling the utilization of a diverse array of PQ public-key signature and encryption algorithms. Moreover, for symmetric cryptography, the adoption of lengthier keys enhances security.

Nevertheless, like any protocol, quantum-resistant S/MIME grapples with certain challenges and vulnerabilities. These encompass potential susceptibilities linked to digital certificates employed in S/MIME, as discussed in Table 11. There is also the lurking danger of private key exposure via side-channel attacks, such as cache side-channel attacks, exploiting observable timing variances and cache access patterns. Spoofing, tampering, repudiation, information disclosure (via side-channel attacks), and denial of service attacks are all salient concerns necessitating vigilant mitigation. To temper these threats, Quantum-resistant S/MIME deploys a spectrum of countermeasures. First and foremost, countermeasures against certificate-based authentication, enumerated in Table 11, are instituted. Secondly, to forestall side-channel attacks, remedies listed in Table 1 are implemented. Furthermore, diverse measures are employed to manage congestion, avert fragmentation, and oversee network overhead to diminish the risk of DoS attacks. This encompasses the utilization of congestion control mechanisms, consideration of message sizes in PQ-resistant protocols, and the provisioning of ample bandwidth and redundancy within the infrastructure.

In summary, quantum-resistant S/MIME furnishes a formidable solution for buttressing email communications within the quantum computing epoch. By replacing conventional cryptographic methods with PQ alternatives and enacting judicious countermeasures, S/MIME adeptly defends against quantum threats while upholding the confidentiality and integrity of electronic mail messages.

### 4.1.11. Final discussion: Adapting application layer security for the quantum era

In the intricate landscape of the TCP/IP model, the Application Layer serves as a cornerstone for enabling secure and efficient data

**Table 2**
Application layer protocols, quantum threats and vulnerabilities.

| TCP/IP Layer | Protocols | Standard/Specification | Main components | Objectives | Recommended cipher suite | Quantum threats |
|---|---|---|---|---|---|---|
| Application | SSH (v2) | RFC 4251 (Ylonen and Lonvick, 2006a), RFC 4252 (Ylonen and Lonvick, 2006), RFC 4253 (Ylonen and Lonvick, 2006), RFC 4254 (Ylonen and Lonvick, 2006), RFC 4256 (Ylonen and pLonvick, 2006), RFC 4335 (Wing, 2006), RFC 4419 (Klyne and Chown, 2006), RFC 5656 (Stebila and Green, 2009), RFC 8308 (Bider et al., 2018) | • Transport Layer Protocol, • SSH User Authentication Protocol, • Connection Protocol, • Additional Components including Secure File Transfer (SFTP), X11 Forwarding for graphical applications, and Port Forwarding for creating secure tunnels. | • Enabling secure access to remote systems and the delivery of other network services across untrusted or insecure networks, • Facilitating robust authentication, data integrity, and data confidentiality to safeguard the integrity and privacy of transmitted information. | • Public-Key Crypto: EdDSA, ECDSA, RSA and DSA, ECDH and DH (broken by Shor's Algo.) • Symmetric Crypto: AES, RC4, 3DES, DES, ChaCha20-Poly1305, SHA1/SHA2, MD5 (Weakened by Grover's Algo.) | • Spoofing, • Tampering, • Info. Disclosure. |
| | SFTP | Friedl and Ylonen (2021) Barrett et al. (2008) | • SSH Protocol for secure communication. • File Transfer Operations, encompassing: (a) file transfer (uploading and downloading) and (b) file Operations (listing directory contents, renaming, deleting, creating directories, changing file permissions). • Error Handling for managing exceptions and issues. | • Facilitating Secure and Encrypted File Access, Transmission, and Efficient File Management, • Strengthening Data Security with Robust Client/Server Authentication, Ensuring Data Integrity, and Maintaining Confidentiality. | • Public-Key Crypto: RSA, DSS, DH (broken by Shor's Algo.) • Symmetric Crypto: 3DES, blowfish, twofish, serpent, IDEA, CAST, AES, HMAC- (MD5, SHA1) (Weakened by Grover's Algo.) | • Spoofing, • Tampering, • Info. Disclosure. |
| | FTPS | RFC 4217 (Ford-Hutchinson and Adams, 2005) | • SSL/TLS Encryption, • Authentication Methods including username and password, public key certificates, and client-side SSL/TLS certificates, • Control and Data Channels, • Firewall and Network Configuration. | • Enhancing the security of File Transfer Protocol (FTP) by implementing Transport Layer Security (TLS) to safeguard data transmissions, • Support connection authentication, integrity and confidentiality. | • Public-Key Crypto: RSA, ECDH (broken by Shor's Algo.) • Symmetric Crypto: AES, SHA2, ChaCha20-Poly1305 (Weakened by Grover's Algo.) | • Spoofing, • Tampering, • Info. Disclosure. |
| | DNSSEC | RFC 4033 (Arends et al., 2005a), RFC 4034 (Arends et al., 2005c), RFC 4035 (Arends et al., 2005b) | • Key Management including the management of Zone Signing Keys (ZSK) and Key Signing Keys (KSK), • Data Verification involving the verification of DNSKEY records and Resource Record Signature (RRSIG) records. • Security Measures encompassing the use of Delegation Signer (DS) Records, NSEC (Next Secure), and NSEC3 (Next Secure version 3) records. | • Bolstering the security of the domain name system by incorporating cryptographic signatures into standard DNS records, thus reinforcing the Internet's ability to withstand a wide range of cyberattacks, • Enhancing the Internet's resilience against diverse cyber threats by providing dependable data origin authentication and preserving the integrity of data transmissions. | • Public-Key Crypto: RSA, DSA, ECDSA (broken by Shor's Algo.) • Symmetric Crypto: SHA1/SHA2/SHA3 (Weakened by Grover's Algo.) | • Spoofing, • Tampering. |
| | SAML (v2) | RFC 7522 (Campbell and Mortimore, 2015) | • Single Sign-On (SSO), • Single Logout (SLO), • Assertion Exchange, • Additional components, such as assertions, bindings, profiles, and metadata. | • Facilitating centralized authentication for multiple applications, • Enabling robust authentication and authorization capabilities across various services and resources. | • Public-Key Crypto: RSA, DSA (broken by Shor's Algo.) | • Spoofing, • Elevation of Privilege. |
| | OAuth (v2) | RFC 6749 (Hardt, 2012), RFC 6750 (Jones and Hardt, 2012), RFC 6819 (Lodderstedt et al., 2013), RFC 7662 (Richer et al., 2015), RFC 7636 (Parecki and Waite, 2015) | • Access Token Acquisition through methods such as the Authorization Code Grant, Resource Owner Password Credentials (ROPC) Grant, and Client Credentials Grant, • Token Management including refreshing and exchanging access tokens, • Security Enhancements encompassing security measures like the Proof Key for Code Exchange (PKCE) Extension and other mechanisms that enhance security during the authorization process. | • Facilitating secure and controlled access for websites or applications to resources hosted by other web applications on behalf of a user, leveraging assertion tokens for this purpose, • Providing a framework for authorization mechanisms, ensuring that access to resources is granted only to authorized entities, • Ensuring the protection of tokens through mechanisms like signature verification to prevent unauthorized access, • Optionally allowing the use of TLS to enhance the secure transmission of tokens. | • Public-Key Crypto: RSA, ECDHE (Use via TLS broken by Shor's Algo.) • Symmetric Crypto: AES, SHA2, ChaCha20- Poly1305 (Used via TLS and weakened by Grover's Algo.), HMAC-SHA1 for Token signature (Weakened by Grover's Algo.) | • Elevation of Privilege. |

**Table 2** (*continued*).

| TCP/IP Layer | Protocols | Standard/Specification | Main components | Objectives | Recommended cipher suite | Quantum threats |
|---|---|---|---|---|---|---|
| | Kerberos (v5) (Neuman et al., 2005) | RFC 4120 (Bradner, 2005), RFC 3961 (Raeburn et al., 2005), RFC 6806 (Hartman et al., 2013), RFC 6113 (Zhu et al., 2011), RFC 8018 (Gutmann, 2017) | • Authentication Service (AS), • Ticket Granting Service (TGS), • Client/Server Authentication, • Additional elements such as Ticket Renewal, Ticket Expiry/Revocation, and Password Change functionality. | • Facilitating secure access authentication to systems over an untrusted network, such as the Internet, • Enabling robust authentication for both clients and servers within client–server applications. | • Public-Key Crypto: - • Symmetric Crypto: HMAC/AES, MD4, MD5, HMAC-SHA1/SHA2/SHA3, CMAC/camellia (Weakened by Grover's Algo.) | • Spoofing. |
| | LDAP (v3) | RFC 4511 (Melrose and Dawson, 2006) | • Simple Authentication and Security Layer (SASL), • Start TLS to upgrade a non-secure LDAP connection to a secure one by establishing a Transport Layer Security (TLS) or Secure Sockets Layer (SSL)-encrypted connection, • LDAP Protocol Extensions including (a) LDAP Controls, (b) Extended Operations, (c) Referral Handling, and (d) Password Policy. | • Facilitating access, authorization, and maintenance of directory services for distributed directory information services across an IP network, • Providing optional support for authentication methods, including: (a) No authentication, (b) Basic authentication, (c) Simple Authentication and Security Layer (SASL), • Optionally ensuring communication confidentiality and data integrity through Transport Layer Security (TLS). | • Public-Key Crypto: optional use of RSA, ECDH via TLS as mentioned in RFC 8446 (Rescorla, 2018a) (broken by Shor's Algo.) • Symmetric Crypto: optional use of AES, SHA2, ChaCha20-Poly1305 via TLS as mentioned in RFC 8446 (Rescorla, 2018a), optional use of CRAM-MD5 via SASL as mentioned in RFC 4422 (Melnikov, 2006; Melnikov and Zeilenga, 2006), (Weakened by Grover's Algo.) | • Spoofing (via credential or SASL), • Tampering (via optional use of TLS), • Info. Disclosure (via optional use of TLS), • Elevation of Privilege. |
| | PGP | RFC 4880 (Callas et al., 2007b), RFC 1991 (Schiller, 1996) | • Key Management including key generation, revocation, and expiration. • Web of Trust to establish the authenticity of public keys. • Other Components including Symmetric/ Asymmetric Encryption, Hashing, Digital Signatures, and Compression. | • Ensuring the secure transmission of data through the use of encrypted emails and files, guaranteeing both privacy and authenticity, • Providing robust protection for sensitive information by adhering to key security principles such as authentication, data integrity, non-repudiation, and confidentiality. | • Public-Key Crypto: RSA, Elgamal, DH, DSA (broken by Shor's Algo.). • Symmetric Crypto: MD5, SHA1, CAST, IDEA, or Triple-DES (Weakened by Grover's Algo.). | • Spoofing, • Tempering, • Repudiation, • Info. Disclosure. |
| | S/MIME (v4) | RFC 8551 (Druta et al., 2019) | • Cryptographic elements including certificate handling, PKI, X.509 Certificates, symmetric/ asymmetric encryption for email protection, message digests, and digital signatures, • Message specification including message formatting and structure. | • Ensuring the privacy and security of electronic messaging, • Facilitating authentication, integrity, non-repudiation, and confidentiality in electronic communication, • Adhering to and complying with diverse data protection and privacy regulations. | • Public-Key Crypto: RSA, DSA, Elliptic Curve (broken by Shor's Algo.) • Symmetric Crypto: AES (Weakened by Grover's Algo.) | • Spoofing, • Tampering, • Repudiation, • Info. Disclosure. |

**Table 3**

Quantum-resistant application layer protocols: Safeguarding against quantum threats.

| TCP/IP Layer | Protocols | Possible quantum-resistant solutions | Challenges and vulnerabilities | Possible threats | Possible countermeasures |
|---|---|---|---|---|---|
| Application | SSH | • OpenSSH: A quantum-resistant version of the widely-used SSH protocol (Author's Name, 2022; Quantum-Safe OpenSSH Team, 2023), • OQS-libssh: An implementation of the SSH protocol that integrates quantum-resistant cryptography through the Open Quantum Safe (OQS) library (Open Quantum Safe (OQS), 2022, 2024). | • Significant communication overhead, • Potential network congestion, • Fragmentation problems leading to retransmissions, • Susceptibility to denial-of-service attacks, • Risk of data exfiltration and information disclosure through side-channel and mathematical analysis attacks, as indicated in Table 1. | • Info. Disclosure, • DoS. | • **To Mitigate Significant Communication Overhead:** Implement congestion control mechanisms to proactively manage and alleviate network congestion issues (Jay et al., 2018; Bohloulzadeh and Rajaei, 2020; Jay et al., 2019). • **To Prevent Fragmentation and Retransmission Issues:** Consider the larger message size introduced by quantum-resistant protocols compared to the classical ones, thus reducing the likelihood of fragmentation and retransmissions (Müller et al., 2020; Beernink, 2022). • **To Address Susceptibility to Denial-of-Service Attacks:** Ensure that the network infrastructure can handle increased communication overhead. This includes providing adequate bandwidth, introducing redundancy into the infrastructure, and deploying dedicated DDoS resilience hardware and software modules such as firewalls, DDoS protection appliances, and properly configuring network hardware to defend against DDoS attacks (Garcia and Blandon, 2022a; Liu et al., 2018). • **To Enhance Resistance Against Side-Channel and Mathematical Analysis Attacks:** Refer to the solutions listed in Table 1 for specific strategies to bolster protection against these types of attacks. |
| | SFTP | • To make the SFTP protocol quantum-resistant, consider replacing the traditional SSH component with quantum-safe alternatives like OpenSSH (Author's Name, 2022; Quantum-Safe OpenSSH Team, 2023) or OQS-libssh (Open Quantum Safe (OQS), 2022). | • Similar Vulnerabilities to quantum-resistant SSH: Quantum-resistant SFTP shares common security concerns with quantum-resistant SSH when transitioning to quantum-resistant protocols. The vulnerabilities and challenges discussed for quantum-resistant SSH are also applicable to quantum-resistant SFTP. | • Info. Disclosure, • DoS. | • **Addressing Common Vulnerabilities Shared with quantum-resistant SSH:** Countermeasures recommended for quantum-resistant SSH are also relevant for quantum-resistant SFTP due to their shared vulnerabilities. These countermeasures include strategies to control network congestion, prevent fragmentation, and enhance resistance against various attacks, all of which are essential during the transition to quantum-resistant protocols. |
| | FTPS | • Possible quantum-resistant solutions for implementing the quantum-resistant FTPS Protocol involve employing OQS-OpenSSL (Open Quantum Safe (OQS), 0000) or KEMTLS (Schwabe, 2023) in place of traditional TLS for secure file transfer. | • Quantum-Resistant FTPS faces the same challenges and vulnerabilities as those related to quantum-resistant TLS, as detailed in Table 5. | • Info. Disclosure, • DoS. | • **Utilizing Shared Countermeasures for Quantum-Resistant TLS:** The countermeasures delineated for quantum-Resistant TLS in Table 5 are equally applicable and effective for enhancing the security of quantum-resistant FTPS. |
| | DNSSEC | • DNSSEC in which (a) classic algorithms are upgrade to PQ ones, (b) longer symmetric keys, with a focus on robust hash algorithms, are used, (c) the size constraints of DNS messages are expand, particularly considering the inclusion of DNSSEC additional information like signatures and keys within the limited 512 bytes (Müller et al., 2020). • Note: In DNSSEC, X.509 certificates are not employed. Instead, zone administrators generate one or more public/private key pairs to create digital signatures, which are utilized to verify the authenticity and integrity of DNS data. | • Significant Communication Overhead, • Network Congestion, • Fragmentation Issues for DNSSEC Records, especially when they need to be transmitted over a single UDP packet or via TCP. This can require higher processing overhead for setting up, tearing down, and downloading larger packets compared to traditional DNSSEC. • Potential Denial of Service (DoS) Attacks, • Information Disclosure via Side-Channel and Mathematical Analysis Attacks as mentioned in Table 1. | • Info. Disclosure, • DoS. | • **Controlling Network Congestion:** Implement congestion control mechanisms, such as those discussed in references like (Jay et al., 2018; Bohloulzadeh and Rajaei, 2020; Jay et al., 2019), to prevent or alleviate network congestion during the operation of quantum-resistant DNSSEC. • **Addressing Fragmentation Issues and Triggering Retransmissions:** Consider the larger message sizes associated with quantum-resistant protocols compared to their classic counterparts, as highlighted in Müller et al. (2020), Beernink (2022), to prevent fragmentation and manage retransmissions effectively. When implementing PQ public-key cryptography instead of classical methods, consider the extreme size of a typical DNSSEC message (up to 1232 bytes). • **Enhancing Tolerance Against Communication Overhead and Preventing DoS:** To counteract the impact of increased communication overhead and mitigate DoS threats, take measures such as providing adequate bandwidth, limiting response rates and sizes, incorporating redundancy into the infrastructure, and deploying DDoS resilience hardware and software modules like ingress filtering and firewalls, as well as adopting DDoS protection appliances and configuring network hardware to resist DDoS attacks (Garcia and Blandon, 2022a; Liu et al., 2018). • **Protecting Against Side-Channel and Mathematical Analysis Attacks:** Refer to the solutions outlined in Table 1 to bolster quantum-resistant DNSSEC's immunity against side-channel and mathematical analysis attacks. |
| | SAML | • SAML, in which: (a) RSA or DSA, two recommended ciphers in its suite, is substituted with PQ cryptographic algorithms. (b) Symmetric cryptography with longer keys is employed to enhance security. (c) Certificates used within the SAML framework are replaced with PQ certificates, as detailed in Section 4.5.5. (d) The optional use of TLS in SAML should be replaced with the possibility of employing quantum-resistant TLS solutions, as referenced in Table 4. This transition is essential to ensure end-to-end security and protection against quantum threats. • Note: SAML does not inherently impose limitations on the size of typical public keys, allowing for the integration of any PQ public-key cryptographic algorithms. | • Significant Communication Overhead, • Network Congestion, • Fragmentation Issues Triggering Retransmission, • Denial of Service Attacks, • Information Disclosure via Side-Channel Attacks, by exploiting observable timing differences and cache access patterns, as indicated in Table 1, • Vulnerabilities Related to Optional TLS Usage, • Vulnerabilities associated with the certificates used in SAML are outlined in Table 11. | • Info. Disclosure, • DoS | • **To Control Congestion:** Employ congestion control mechanisms to prevent or alleviate network congestion issues, drawing from established strategies and solutions in the field (Jay et al., 2018; Bohloulzadeh and Rajaei, 2020; Jay et al., 2019). • **To Avoid Fragmentation and Trigger Retransmission:** Optimize message size, utilize an appropriate transport protocol like HTTPS, employ compression techniques, ensure proper configuration of network infrastructure, and engage in regular testing and monitoring to mitigate fragmentation-related problems. • **To Increase Tolerance Against Communication Overhead and Prevent DoS:** Provide the necessary bandwidth, implement redundancy in the infrastructure, deploy DDoS resilience hardware and software modules such as firewalls, adopt DDoS protection appliances, and configure network hardware to withstand DDoS attacks (Garcia and Blandon, 2022a; Liu et al., 2018). • **To Mitigate Side-Channel Attacks:** Refer to the solutions outlined in Table 1 to enhance immunity against side-channel attacks. • **Leverage Countermeasures for TLS and SAML Certificates:** Utilize the countermeasures mentioned for TLS protocol vulnerabilities and those outlined in Table 11 for certificate-related issues, as these solutions are also applicable to Quantum-Resistant SAML. |

**Table 3** (continued).

| TCP/IP Layer | Protocols | Possible quantum-resistant solutions | Challenges and vulnerabilities | Possible threats | Possible countermeasures |
|---|---|---|---|---|---|
| | OAuth | • OAuth (v2) can be made quantum-resistant by replacing the existing token signatures, such as HMAC-SHA1, with post-quantum (PQ) alternatives. This transition requires the use of longer cryptographic keys, while taking into account the maximum allowable access token length of 2048 bytes. It is important to note that OAuth (v2) does not specify a cryptographic mechanism, except for the optional use of TLS, as detailed in RFC 6819 and RFC 6749. In a quantum-resistant context, it is advisable to replace traditional TLS with quantum-resistant TLS to enhance the security of OAuth (v2), as elaborated in Table 11. | • Vulnerabilities tied to the optional use of TLS have been previously addressed in the discussion regarding the TLS protocol. | • Info. Disclosure, • DoS. | • **Utilizing Shared Countermeasures for Quantum-Resistant TLS:** The countermeasures outlined for the TLS protocol vulnerabilities are equally applicable to address and mitigate the vulnerabilities related to the optional use of TLS in Quantum-Resistant OAuth. |
| | Kerberos | • Utilizing Kerberos with longer symmetric keys, substantially increases the computational effort needed to break them using quantum algorithms. | • Symmetric Key Vulnerabilities: The use of symmetric keys derived from passwords can introduce vulnerabilities, as these keys might not be fully random and could be susceptible to attacks such as Birthday attacks and Grover's algorithm (Grover, 2023). | • Spoofing. | • **Implementing Strongly Random Long-Term Symmetric Keys:** To address this vulnerability, it is advisable to generate and employ strongly random long-term symmetric keys that are not derived from passwords. These keys should be unique for each user, and possibly each device, and should be securely distributed and installed to enhance security. |
| | LDAP | • In the context of LDAP, quantum-resistant solutions may entail the following measures: (a) Employing longer and more secure passwords or symmetric keys in the SASL (Simple Authentication and Security Layer) mechanism to bolster resistance against birthday attacks. (b) Substituting the optional use of TLS (Transport Layer Security) with quantum-resistant alternatives like OQS-OpenSSL (Open Quantum Safe (OQS), 0000) or KEMTLS (Schwabe, 2023). (c) Utilizing quantum-resistant certificates in place of traditional ones, as detailed in Section 4.5. | • Insecure Use of Passwords or Symmetric Keys in SASL: The optional use of a password or symmetric key in SASL for authentication may lack sufficient randomness, rendering it vulnerable to attacks like the Birthday attack and Grover's algorithm (Grover, 2023). <br> • Shared Vulnerabilities with Quantum-Resistant TLS: As LDAP allows for optional use of TLS, it inherits the same vulnerabilities as Quantum-Resistant TLS, including those associated with potential communication overhead, network congestion, fragmentation issues leading to retransmissions, susceptibility to denial-of-service attacks, and the risk of data exfiltration and information disclosure through side-channel and mathematical analysis attacks, as indicated in Table 5. | • Spoofing, • Info. Disclosure, • DoS. | • **Implementing Strong, Random Long-term Symmetric Keys or Passwords:** To address the vulnerability in SASL, it is crucial to provide each user with a strongly random long-term symmetric key or password that is securely distributed and installed, enhancing the security of authentication methods. <br> • **Leveraging Countermeasures from quantum-resistant TLS:** Given the shared vulnerabilities with Quantum-Resistant TLS, it is advisable to apply the countermeasures recommended for Quantum-Resistant TLS. These countermeasures encompass strategies to manage network congestion, prevent fragmentation, and enhance resistance against various attacks, which are equally relevant in the context of LDAP security. |
| | PGP | • OpenPGP (Callas et al., 2007a) can achieve quantum resistance through the following measures: (a) Replacing traditional public-key cryptography with post-quantum (PQ) cryptographic methods. (b) Increasing key lengths for symmetric cryptography. (c) Note: In contrast to the conventional use of 4096-bit public keys in OpenPGP, the integration of PQ public keys may require longer key lengths for symmetric cryptography. This adjustment is essential to uphold robust security in the quantum computing era. Longer key lengths for symmetric cryptography effectively mitigate potential vulnerabilities and challenges arising from advances in quantum computing, ensuring the ongoing security of OpenPGP in the face of quantum threats. | • Chain of Trust-Induced Challenges: Quantum-Resistant PGP relies on a chain of trust, leading to: Substantial communication overhead. Risk of network congestion. Fragmentation issues. Susceptibility to denial-of-service attacks. Private Key Retrieval through Side-Channel Attacks: Private keys can be vulnerable to retrieval through side-channel attacks, using techniques detailed in Table 1. | • Spoofing, Tampering, Repudiation, and Info. Disclosure (via side-channel attacks), • DoS. | • **Mitigating Side-Channel Attacks:** To safeguard against side-channel attacks, reference the solutions provided in Table 1 for effective mitigation. <br> • **Controlling Congestion:** Employ congestion control mechanisms to manage and alleviate network congestion (e.g., (Jay et al., 2018; Bohloulzadeh and Rajaei, 2020; Jay et al., 2019)). <br> • **Preventing Fragmentation and Triggering Retransmission:** Consider the enlarged size of a typical message when using quantum-resistant protocols instead of classical ones (e.g., (Müller et al., 2020; Beernink, 2022)). <br> • **Enhancing Tolerance to Communication Overhead and DoS Attacks:** To bolster resistance against substantial communication overhead and to mitigate denial-of-service threats, consider the following measures: (a) Provision of adequate bandwidth, (b) Imposing limits on response rates and response sizes, (c) Incorporating redundancy into the infrastructure. (d) Deployment of DDoS resilience hardware and software modules, such as ingress filtering and firewalls, (e) Adoption of DDoS protection appliances, (f) Configuring network hardware to withstand DDoS attacks (e.g., (Garcia and Blandon, 2022a; Liu et al., 2018)). <br> • **Mitigating Side-Channel Attacks:** For additional protection against side-channel attacks, refer to the solutions listed in Table 1. |
| | S/MIME | • S/MIME can achieve quantum resistance through the following measures: (a) Replace classic public-key cryptography with post-quantum (PQ) cryptography in S/MIME. (b) Employ symmetric cryptography with longer keys. (c) Use PQ certificates, as elaborated in Section 4.5, to enhance security. • Note: In the context of S/MIME, there are no constraints on the size of public keys, allowing for the use of any PQ public-key signing and encryption algorithms (as discussed in Schaad et al. (2019)). | • Reliance on Digital Certificates: Quantum-Resistant S/MIME relies on digital certificates for user identification, which introduces specific vulnerabilities outlined in Table 11. <br> • Private Key Retrieval via Side-Channel Attacks: There is a potential risk of adversaries gaining access to private keys through side-channel attacks, such as cache side-channel attacks, as described in Table 1. | • Spoofing, Tampering, Repudiation, and Info. Disclosure (via side-channel attacks), • DoS. | • **Addressing Certificate-Based Authentication Vulnerabilities:** To mitigate the vulnerabilities associated with certificate-based authentication in Quantum-Resistant S/MIME, refer to the solutions detailed in Table 11. <br> • **Immunity Against Side-Channel Attacks:** To safeguard against the risk of private key retrieval through side-channel attacks, implement the protective measures specified in Table 1. |

exchange among software applications. Serving as the gateway to network services and user interfaces, this layer plays an indispensable role in modern network communication. In Table 2, we have meticulously examined a variety of network and security protocols residing within the Application Layer, delving into their standardization, core components, primary functions, recommended cipher suites, and their current susceptibility to quantum-related threats.

Moving forward, our exploration continues with Table 3, which illuminates the path towards quantum-resistant alternatives for these protocols within the TCP/IP model. This section sheds light on the inherent challenges and potential vulnerabilities associated with the adoption of these advanced solutions. Additionally, it outlines the emerging quantum threats that these quantum-resistant alternatives may confront and the countermeasures that have been devised to effectively neutralize these threats.

In summary, our deep dive into the Application Layer underscores the imperative need to embrace quantum-resistant solutions, especially in the face of rapid advancements in quantum computing capabilities. While these alternatives do introduce their own complexities and potential vulnerabilities, confronting the impending quantum threat is non-negotiable.

The implementation of strategic countermeasures, optimization of communication overhead, and the fortification of infrastructure resilience stand as pivotal actions to safeguard the security and dependability of Application Layer protocols in this quantum-empowered era. Organizations must maintain vigilance and adaptability in their security measures to proactively address the ever-evolving landscape of threats.

### 4.2. Transport layer

The Transport Layer plays a crucial role in enabling end-to-end communication in network systems. It includes important protocols such as TCP (Transmission Control Protocol) and UDP (User Datagram Protocol), which form the foundation of Internet data transmission. This layer is equipped with security mechanisms like TLS (Transport Layer Security) and mTLS (Mutual Transport Layer Security), which ensure the confidentiality and integrity of data, preventing unauthorized access and tampering. Nevertheless, the emergence of quantum computing presents a significant security concern due to its potential to compromise the encryption methods that TLS and mTLS rely on. Therefore, it is essential to develop security measures that are resistant to quantum threats for the Transport Layer to effectively address this impending risk.

#### 4.2.1. TLS protocol

Transport Layer Security (TLS), a critical protocol in modern network communication, shares several commonalities with SSH (v2). It is presented in multiple RFCs (Dierks and Allen, 1999; Dierks and Rescorla, 2006; Dierks and Rescorla, 2008; Rescorla, 2018b; Sheffer et al., 2015; Rescorla and Modadugu, 2018; Fraser and Cooper, 2015). Both TLS and SSH serve the central purpose of ensuring secure data transmission over potentially insecure networks, accomplishing this by providing authentication, data integrity, and confidentiality. In TLS, this security is achieved through a combination of cryptographic techniques, including public-key cryptography (such as RSA and ECDH) and symmetric cryptography (like AES and SHA2). It is worth noting that these cryptographic methods may become vulnerable to future advancements in quantum computing, such as Shor's and Grover's algorithms.

TLS is a comprehensive approach to network security and is the go-to choice for securing various communication channels, ranging from web browsing to email transmission. It guarantees the confidentiality and integrity of sensitive data. However, like SSH, TLS faces several quality control threats, such as spoofing, tampering, and information disclosure, potentially exploited by a quantum attacker. To maintain

TLS's robustness in the face of evolving threats, it is crucial to remain vigilant and adapt cryptographic strategies as quantum computing technologies advance (Rescorla, 2018a).

To ensure the robustness of TLS in a post-quantum world, post-quantum TLS, also known as Quantum-Resistant TLS, plays a vital role in securing communications. When migrating to post-quantum TLS, several key considerations and challenges must be addressed. One primary challenge is the integration of post-quantum cryptographic solutions into existing TLS protocols. Post-quantum TLS solutions, such as OQS-OpenSSL (Open Quantum Safe (OQS), 0000) and KEMTLS (Schwabe, 2023), replace vulnerable traditional cryptographic algorithms with quantum-resistant alternatives. However, this transition can introduce several issues.

Firstly, the migration to post-quantum TLS can result in a significant increase in communication overhead. Post-quantum cryptographic algorithms often require larger key sizes and more complex operations, leading to larger message sizes and increased processing time. To mitigate this challenge, it is essential to provide sufficient network bandwidth and employ optimization techniques to manage the increased overhead. Secondly, network congestion can become problematic during the migration to post-quantum TLS. Employing congestion control mechanisms is crucial to prevent or alleviate congestion, ensuring smooth communication even with increased data volume.

Fragmentation issues that trigger retransmissions are another concern. The larger message sizes when using post-quantum TLS can lead to fragmentation, potentially causing problems and delays. Properly configuring the network infrastructure to handle these issues is vital. Furthermore, the migration can expose new attack vectors, including the potential for DoS threats. To defend against these threats, network infrastructure should incorporate redundancy, deploy DDoS resilience hardware and software modules, and configure network hardware effectively to mitigate DDoS attacks. Additionally, there is a risk of information disclosure during the migration to post-quantum TLS. Side-channel attacks and mathematical analysis attacks can be used to reveal sensitive information. Mitigating these threats requires implementing the countermeasures provided by post-quantum cryptographic solutions.

In summary, quantum-resistant TLS involves transitioning from vulnerable cryptographic algorithms to quantum-resistant alternatives, addressing the quantum threat but introducing challenges related to communication overhead, network congestion, fragmentation, DoS attacks, and information disclosure. Effective countermeasures, network optimizations, and congestion control mechanisms are crucial to ensuring the security and reliability of quantum-resistant TLS in a post-quantum world (see Table 5).

#### 4.2.2. mTLS protocol

Mutual Transport Layer Security (mTLS) is a robust protocol encompassing various sub-protocols like TLS Handshake, Alert, Change Cipher Spec, and the TLS Record Protocol (Dierks and Rescorla, 2008; Rescorla, 2018b). Its primary goal is to establish a secure communication framework, encompassing authentication, data integrity, and confidentiality. mTLS achieves this through a combination of cryptographic techniques, including public-key cryptography such as RSA and ECDH (though ECDH is susceptible to Shor's algorithm) for key exchange and authentication. Additionally, symmetric ciphers like AES and SHA2 are used for data encryption, with the option to enhance security using ChaCha20-Poly1305 (though vulnerable to Grover's algorithm).

A significant concern regarding quantum attacks on mTLS is the potential for spoofing, where malicious entities attempt to impersonate legitimate parties during communication. It also addresses tampering, involving unauthorized data alterations, and safeguards against information disclosure to ensure sensitive data confidentiality. In the era of quantum computing threats, post-quantum mTLS becomes critical, relying on quantum-resistant cryptographic protocols to defend against potential quantum computer threats. The primary quantum-resistant cryptographic solutions for mTLS include mutual use of OQS-OpenSSL (Open Quantum Safe (OQS), 0000) or KEMTLS (Schwabe, 2023), which both parties involved in communication can adopt.

**Table 4**
Transport layer protocols, quantum threats and vulnerabilities.

| TCP/IP Layer | Protocols | Standard/Specification | Main components | Objectives | Recommended cipher suite | Quantum threats |
|---|---|---|---|---|---|---|
| Transport | TLS (v1.3) | RFC 2246 (Dierks and Allen, 1999), RFC 4346 (Dierks and Rescorla, 2006), RFC 5246 (Dierks and Rescorla, 2008), RFC 8446 (Rescorla, 2018b), RFC 7525 (Sheffer et al., 2015), RFC 8447 (Rescorla and Modadugu, 2018), RFC 7627 (Fraser and Cooper, 2015). | • Handshake Protocol, <br>• Record Protocol, <br>• Alert Protocol, <br>• Change Cipher Spec Protocol, <br>• Higher-layer Protocols enabling secure communication for applications such as HTTP (HTTPS), FTPS, <br>• Other components including encryption algorithms, cryptographic certificates, and additional elements that enhance the security and functionality of TLS. | • Ensuring robust communication security (Rescorla, 2018a). <br>• Facilitating authentication, maintaining data integrity, and preserving confidentiality in data transmission. | • Public-Key Crypto: RSA, ECDH (broken by Shor's Algo.) <br>• Symmetric Crypto: AES, SHA2, ChaCha20-Poly1305 (Weakened by Grover's Algo.) | • Spoofing, <br>• Tampering, <br>• Info. Disclosure. |
| | mTLS | RFC 5246 (Dierks and Rescorla, 2008), RFC 8446 (Rescorla, 2018b) | • The key components of mTLS (Mutual Transport Layer Security) closely mirror those of TLS (Transport Layer Security). | • Ensuring robust communication security. <br>• Enabling authentication, data integrity, and confidentiality support. | • Public-Key Crypto: RSA, ECDH (broken by Shor's Algo.) <br>• Symmetric Crypto: AES, SHA2, ChaCha20-Poly1305 (Weakened by Grover's Algo.) | • Spoofing, <br>• Tampering, <br>• Info. Disclosure. |

**Table 5**
Quantum-resistant transport layer protocols: Safeguarding against quantum threats.

| TCP/IP Layer | Protocols | Possible quantum-resistant solutions | Challenges and vulnerabilities | Possible threats | Possible countermeasures |
|---|---|---|---|---|---|
| Transport | TLS | • OQS-OpenSSL: This solution, extensively described in references like (Open Quantum Safe (OQS), 0000), adheres to the Open Quantum Safe (OQS) framework designed for TLS. <br> • KEMTLS: As outlined in Schwabe (2023), KEMTLS offers an alternative, robust quantum-resistant solution for TLS. | • Shared Vulnerabilities with Quantum-Resistant SSH: Quantum-Resistant TLS encounters similar vulnerabilities to those observed in Quantum-Resistant SSH, including potential communication overhead, network congestion, fragmentation issues leading to retransmissions, susceptibility to denial-of-service attacks, and the risk of data exfiltration and information disclosure through side-channel and mathematical analysis attacks. For a detailed comparison, please refer to Table 1. <br> • Reliance on Digital Certificates: Quantum-Resistant TLS heavily depends on digital certificates for server and client authentication. However, digital certificates have their own inherent drawbacks, which are outlined in Table 11. | • Info. Disclosure, <br> • DoS. | • **Addressing Common Vulnerabilities Shared with Quantum-Resistant SSH:** Given the analogous challenges faced by Quantum-Resistant TLS and Quantum-Resistant SSH, it is prudent to consider countermeasures that have been recommended for Quantum-Resistant SSH. These countermeasures include strategies to mitigate network congestion, prevent fragmentation issues, and enhance resistance against various types of attacks. These efforts are pivotal during the transition to quantum-resistant protocols. <br> • **Tackling Concerns Related to Certificate-Based Authentication:** To address the concerns related to certificate-based authentication in Quantum-Resistant TLS, it is advisable to refer to the solutions outlined in Table 11. These solutions can offer valuable insights into improving the security and reliability of certificate-based authentication mechanisms in a quantum-resistant context. |
| | mTLS | • Simultaneous Adoption of OQS-OpenSSL: Both communicating parties can opt for OQS-OpenSSL (Open Quantum Safe (OQS), 0000) for quantum-resistant mTLS. <br> • Reciprocal Utilization of KEMTLS: Both parties can employ KEMTLS (Schwabe, 2023) for quantum-resistant mTLS, ensuring mutual protection against quantum threats. | • Quantum-Resistant mTLS inherits all the vulnerabilities and challenges previously discussed for TLS. These include the full spectrum of attacks applicable to TLS, which are equally relevant and concerning in the context of quantum-resistant mTLS. | • Info. Disclosure, <br> • DoS. | • **Leveraging TLS Countermeasures:** The countermeasures discussed for quantum-resistant TLS are equally applicable to quantum-resistant mTLS. By implementing these countermeasures, organizations can enhance the security of their quantum-resistant mTLS implementations and mitigate potential threats. |

However, adopting post-quantum cryptographic protocols does not eliminate all security challenges. Similar to TLS and SSH, quantum-resistant cryptographic solutions for mTLS face risks like communication overhead, network congestion, fragmentation leading to retransmission, and the potential for DoS attacks. Attackers may also attempt side-channel attacks or exploit mathematical vulnerabilities, as detailed in Table 1. To counter these challenges and protect mTLS communications, several measures can be employed. These include implementing congestion control mechanisms, avoiding fragmentation and trigger-based retransmissions by considering message size, providing sufficient bandwidth to handle communication overhead, introducing redundancy into infrastructure, and deploying hardware/software modules for DDoS resilience and protection. Additionally, mitigating the risk of side-channel attacks and mathematical analysis vulnerabilities requires referring to the solutions outlined in Table 1. These solutions involve secure coding practices, using encryption techniques less susceptible to side-channel attacks, and regularly updating cryptographic algorithms to address emerging threats.

In summary, quantum-resistant mTLS plays a critical role in securing communications in a post-quantum computing landscape. By adopting quantum-resistant cryptographic protocols like OQS-OpenSSL and KEMTLS, along with implementing appropriate countermeasures against potential threats, organizations can ensure the confidentiality, integrity, and availability of their sensitive data and communications in the face of evolving cybersecurity challenges.

### 4.2.3. Final discussion: Adapting transport layer security for the quantum era

In the face of quantum threats and vulnerabilities, the importance of securing network communication at Application Layer, specifically through Transport Layer Security (TLS) and Mutual Transport Layer Security (mTLS) protocols, cannot be overstated. The widely adopted TLS v1.3 protocol, although effective in many ways, is susceptible to potential quantum threats due to vulnerabilities in its cryptographic algorithms. As a response to this emerging challenge, quantum-resistant TLS and mTLS protocols are under consideration to fortify network security. These quantum-resistant protocols, such as OQS-OpenSSL and KEMTLS, aim to replace the vulnerable cryptographic algorithms used in traditional TLS and mTLS with quantum-resistant alternatives. However, the adoption of these solutions is not without its challenges. Some of these challenges include an increase in communication overhead, network congestion, fragmentation issues, and the potential for DoS attacks. To mitigate these challenges and ensure the resilience of these quantum-resistant protocols, various countermeasures need to be implemented.

Optimization techniques, congestion control mechanisms, and appropriate network infrastructure configuration are essential to address communication overhead and network congestion. Redundancy in infrastructure and the deployment of hardware and software modules for DDoS resilience play a crucial role in protecting against potential DoS attacks. Furthermore, as with any security protocol, quantum-resistant TLS and mTLS may still face threats, including the risk of information disclosure. Side-channel attacks and mathematical vulnerabilities remain potential attack vectors. To combat these threats effectively, it is imperative to implement the countermeasures provided by quantum-resistant cryptographic solutions.

In summary, quantum threats from emerging quantum computing technologies necessitate the adaptation of Transport Layer security protocols. Quantum-resistant TLS and mTLS protocols provide a vital defense against these threats, ensuring the confidentiality, integrity, and availability of sensitive data and communications. While their adoption comes with challenges, the proactive implementation of countermeasures and optimization strategies is essential for a secure transition. The accompanying tables, Table 4 and Table 5, offer a comprehensive analysis of various Transport Layer network and security protocols. They provide insights into their standardization, essential components, primary purposes, and recommended cipher suites, and the quantum-related threats they may currently face. Furthermore, Table 5 delves into the potential quantum-resistant alternatives for these protocols. It highlights the challenges and vulnerabilities associated with their adoption and outlines the countermeasures designed to mitigate these threats, ensuring a secure quantum-era network environment.

### 4.3. Internet layer

The Internet Layer, a fundamental component of the TCP/IP model, is pivotal in data transmission, encompassing critical functions such as routing, logical addressing, and Wide Area Network (WAN) delivery. Its primary responsibility is routing and forwarding data packets across the network, involving tasks related to logical addressing, like IP addresses. This layer plays a crucial role in determining the most efficient path for data to travel from the source to the destination, ensuring data reaches the correct destination irrespective of the physical network layout. Efficient data packet delivery relies on the capabilities of this layer. Moreover, the Internet Layer hosts essential security protocols that bolster data transmission security. Notably, Internet Key Exchange (IKE) and IPsec (Internet Protocol Security) primarily operate within this layer to fortify data transmission. IKE is a pivotal protocol for establishing secure connections and managing encryption keys in IPsec-based virtual private networks (VPNs). It facilitates secure negotiation and exchange of encryption keys between communicating devices, safeguarding data confidentiality and integrity during transit. It is crucial to recognize that while IKE and IPsec are fundamental for secure data transmission, the advent of quantum computing poses a substantial threat to widely-used public-key encryption algorithms, including those employed by IKE and IPsec. Hence, there exists an imperative need to embrace quantum-resistant alternatives to uphold data transmission security in the face of evolving technological threats.

### 4.3.1. IKE protocol

One of the primary quantum-resistant security protocols for the Internet Layer is IKE. IKE is a critical protocol used for establishing Security Associations (SA), negotiating security policies, and managing cryptographic keys in IPsec VPNs. Its primary purpose is to ensure the confidentiality, integrity, and authenticity of data exchanged over a network. Classic IKE relies on both public-key cryptography, specifically the Diffie–Hellman (DH) key exchange, and symmetric cryptography, including AES and HMAC-SHA algorithms, for securing communications (Kaufman and Eronen, 2014). However, IKE is not immune to the emerging threat posed by quantum computing. Quantum computers, such as those leveraging Shor's algorithm, have the potential to break the security provided by classic IKE's DH key exchange. This vulnerability could lead to quantum attackers easily intercepting and decrypting confidential information, posing a grave risk to the confidentiality and security of network communications. In addition to quantum threats, IKE is also susceptible to common cryptographic threats such as spoofing, tampering, and information disclosure. Therefore, it is crucial for organizations to consider post-quantum cryptography and other security measures to safeguard their IKE-based VPNs against both conventional and quantum-based attacks. Upgrading to quantum-resistant cryptographic algorithms and enhancing overall security practices will be essential to mitigate these risks and ensure the continued effectiveness of IKE in a quantum computing era.

Quantum-resistant IKE is a critical component in securing communications against the looming threat of quantum computing. In the context of IKE, several possible solutions have been proposed to withstand quantum attacks. One of the prominent solutions is PQ IKE, as outlined in RFC 8784 titled "Mixing Preshared Keys in the IKE (v2) for Post-quantum Security" (Kivinen et al., 2020). However, implementing quantum-resistant IKE comes with its own set of challenges and potential vulnerabilities. One concern is the generation of post-quantum preshared keys from passwords, which may have low entropy and lack full randomness, making them vulnerable to attacks such as Birthday attacks and Grover's algorithm. Additionally, there is a risk of information disclosure through side-channel attacks, leveraging observable timing differences and cache access patterns. Moreover, PQ IKE relies on digital certificates, which can introduce vulnerabilities, as discussed in Table 11. These issues include issues such as certificate-based attacks and out-of-bounds memory access vulnerabilities.

To address these challenges and enhance the security of quantum-resistant IKE, various countermeasures can be implemented. It is crucial to ensure that post-quantum preshared keys have sufficient entropy and randomness to withstand attacks. Additionally, measures must be taken to protect against side-channel attacks by applying techniques mentioned in Table 1. Furthermore, countermeasures for certificate-related vulnerabilities are essential. These measures include updating PQ certificates as per the recommendations outlined in Section 4.5 and adopting solutions provided in Table 11 to mitigate certificate-based attacks. In summary, quantum-resistant IKE, as outlined in RFC 8784 (Kivinen et al., 2020), is a vital step towards securing communication in a post-quantum computing era. However, its successful implementation requires careful consideration of key generation, vulnerability to side-channel attacks, and the use of digital certificates, along with the implementation of appropriate countermeasures to address these challenges effectively.

### 4.3.2. IPsec protocol

IPsec is a critical suite of protocols designed to secure communication over IP networks, providing essential security and authentication. IPsec encompasses various sub-protocols, including Information Key Exchange (IKE), Authentication Header (AH), and Encapsulating Security Protocol (ESP), each serving distinct roles (Kent and Seo, 2005). Its primary objectives are to establish secure, authenticated, and reliable communication, ensuring data origin authentication, connectionless integrity, and confidentiality. However, like many security protocols relying on cryptographic algorithms, IPsec faces evolving threats from quantum computing. Quantum attackers armed with quantum computers can potentially compromise the cryptographic primitives used in IPsec. For instance, Shor's algorithm can undermine public-key cryptography algorithms like DH, ECDH, RSA, ECDSA, and DSA, jeopardizing IPsec's confidentiality and authenticity guarantees. Additionally, Grover's algorithm may weaken symmetric cryptography, including AES, HMAC-SHA1/SHA2, 3DES, MD5, and ChaCha20-Poly1305, making IPsec susceptible to threats like spoofing, tampering, and information disclosure.

As quantum computing continues to advance, the adaptation of IPsec protocols with post-quantum cryptography becomes essential to mitigate quantum threats. Noteworthy quantum-resistant IPsec protocols under consideration for this purpose include OpenVPN (Microsoft, 2022), StrongSwan (The strongSwan Team, 2022), and WireGuard (Hülsing et al., 2021). OpenVPN, a widely adopted VPN protocol known for its compatibility advantages, has been a candidate for quantum-resistant adaptation. However, it is important to note that OpenVPN tends to have a longer initiation time and may exhibit potentially lower throughput when compared to alternative solutions. StrongSwan, while rooted in the IPsec family, stands out for its low-latency performance. It is currently being evaluated in the context of quantum resistance, but potential trade-offs should be carefully assessed. WireGuard, a distinct and efficient VPN protocol, has gained recognition for its robust security features. However, it is vital to understand that WireGuard operates independently and is not part of the IPsec standardization process. Its unique cryptographic approach may or may not necessitate quantum-resistant enhancements, depending on future quantum threats. In terms of performance, OpenVPN exhibits a longer initiation time compared to both StrongSwan. On the other hand, it offers better compatibility but might have lower throughput. WireGuard, in contrast, is recognized for its efficiency and is particularly well-suited for bypassing censorship (Dekker and Spaans, 2020). It boasts a shorter initiation time and lower latency compared to the other two protocols. Careful consideration of the specific use case and security requirements is essential when selecting the most suitable IPsec protocol for quantum-resistant communication. Each of these quantum-resistant IPsec protocols has its strengths and weaknesses, including differences in latency, throughput, and compatibility. These IPsec alternatives support both pre-shared keys and digital certificates for securing communications. However, vulnerabilities exist, especially when pre-shared keys are derived from passwords or digital certificates are used. Post-quantum pre-shared keys derived from passwords might have low entropy, lack randomness, and can be broken by a Birthday attack and Grover's algorithm (Grover, 2023). Pre-shared keys can also be recovered via side-channel attacks, as a result of observable timing differences and cache access patterns. Post-quantum digital certificates have drawbacks mentioned in Table 11.

To mitigate these challenges when transitioning to quantum-resistant IPsec protocols, several countermeasures can be implemented. To enhance the security of IPsec in a post-quantum environment, it is crucial to provide strongly random, long-term post-quantum pre-shared keys with sufficient entropy and employ countermeasures against side-channel attacks as described in the solutions presented in the table.

To control congestion, congestion control mechanisms can be employed to manage and alleviate congestion issues. Recognizing the increased message size when using quantum-resistant protocols can help maintain smoother communication and prevent fragmentation and retransmission problems. Ensuring sufficient bandwidth and redundancy and deploying DDoS resilience hardware and software modules like firewalls and DDoS protection appliances can boost tolerance against communication overhead and prevent DoS attacks.

To address the risk of information disclosure and DoS attacks through side-channel or mathematical analysis, refer to the solutions presented in Table 1 to enhance the security of IPsec connections. Bolstering IPsec security in a post-quantum environment, it is critical to use robust, random, and enduring pre-shared keys with ample entropy to withstand quantum-deciphering efforts. Implement defenses against side-channel assaults as delineated in Table 1. Additionally, address the limitations of digital certificates with the countermeasures specified in Table 11. Ultimately, quantum-resistant IPsec aims to mitigate threats like information disclosure and denial of services in a world where quantum computing poses new challenges to cryptographic security.

### 4.3.3. Final discussion: Adapting internet layer security for the quantum era

In this discussion, we have explored the crucial role of Internet Layer protocols, specifically IKE and IPsec, in securing data transmission within modern networks. However, it is evident that these protocols are not impervious to the emerging threats posed by quantum computing, which has the potential to compromise their security mechanisms through advanced algorithms like Shor's and Grover's.

Table 6 provided a comprehensive analysis of these Internet Layer protocols, including their standardization, core components, primary purposes, recommended cipher suites, and the quantum-related challenges they face. It underscores the pressing need for the adaptation of security measures to withstand the quantum era. Additionally, Table 7 presents potential quantum-resistant alternatives, offering insights into the challenges and vulnerabilities associated with their adoption and outlining the potential threats these quantum-resistant solutions may face. This emphasizes the importance of proactively preparing for the quantum threat landscape.

Quantum computing is advancing rapidly, making it imperative for organizations to address quantum threats within the Internet Layer.

**Table 6**
Internet layer protocols, quantum threats and vulnerabilities.

| TCP/IP Layer | Protocols | Standard/Specification | Main components | Objectives | Recommended cipher suite | Quantum threats |
|---|---|---|---|---|---|---|
| Internet | IKE (v2) | RFC 7296 (Kaufman and Eronen, 2014) | • IKE_SA Establishment, • CHILD_SA Establishment, • Other components including (a) rekeying, (b) mobility and multihoming, and (c) Dead Peer Detection (DPD). | • Facilitating Security Association (SA) establishment, policy negotiation, and robust key management for secure communication. • Enabling secure key exchange, Extensible Authentication Protocol (EAP) authentication, ensuring data integrity, and maintaining confidentiality. | • Public-Key Crypto: DH (broken by Shor's Algo.) • Symmetric Crypto: AES, HMAC-SHA1/SHA2, 3DES, MD5, ChaCha20-Poly1305 (Weakened by Grover's Algo.) | • Spoofing, • Tampering, • Info. Disclosure. |
| | IPsec | RFC 4301 (Kent and Seo, 2005) | • IKE (Internet Key Exchange), • AH (Authentication Header), • ESP (Encapsulating Security Protocol), • Security Associations (SA), • Security Policies, • Key Management. | • Ensuring Secure and Authenticated Communication, • Providing Data Origin Authentication, • Ensuring Connection-Less Integrity, • Maintaining Confidentiality. | • Public-Key Crypto: DH, ECDH, RSA. ECDSA (broken by Shor's Algo.) • Symmetric Crypto: AES, HMAC-SHA1/SHA2, 3DES, MD5, ChaCha20-Poly1305 (Weakened by Grover's Algo.) , PSK. | • Spoofing, • Tampering, • Info. Disclosure. |

**Table 7**
Quantum-resistant internet layer protocols: Safeguarding against quantum threats.

| TCP/IP Layer | Protocols | Possible quantum-resistant solutions | Challenges and vulnerabilities | Possible threats | Possible countermeasures |
|---|---|---|---|---|---|
| Internet | IKE | • Quantum-Resistant IKE protocol, as detailed in RFC 8784: "Mixing Preshared Keys in the IKE (v2) for Post-Quantum Security" (Kivinen et al., 2020; Fluhrer et al., 2020). This document outlines a strategy for enhancing the quantum resistance of IKE by incorporating pre-shared keys within the context of IKE (v2). | • Low Entropy Post-Quantum Preshared Keys: Preshared keys derived from passwords in a post-quantum setting may exhibit low entropy, lack full randomness, and be susceptible to attacks such as Birthday attacks and Grover's algorithm (Grover, 2023). • Information Disclosure through Side-Channel Attacks: Quantum-Resistant IKE may be vulnerable to information disclosure, as attackers can potentially recover preshared keys through side-channel attacks. This vulnerability arises from observable timing differences and cache access patterns, as indicated in Table 1. • Dependency on Digital Certificates: Quantum-Resistant IKE relies on digital certificates, which bring their own set of vulnerabilities, as described in Table 11. | • Info. Disclosure, • DoS | • **Generation of Strongly Random, Long-Term Post-Quantum Preshared Keys:** To mitigate the vulnerability of low entropy keys, it is crucial to generate post-quantum preshared keys with a high degree of randomness and sufficient entropy. This will make them more resilient to attacks. • **Immunity against Side-Channel Attacks:** Implement countermeasures to protect against side-channel attacks. Refer to the solutions detailed in Table 1 to fortify the system against information disclosure vulnerabilities stemming from observable timing differences and cache access patterns. • **Addressing Certificate-Related Drawbacks:** To tackle the vulnerabilities associated with the use of digital certificates in Quantum-Resistant IKE, consult the remedies specified in Table 11 to enhance the security of certificate-based authentication. |
| | IPsec | • OpenVPN (Microsoft, 2022), • StrongSwan (The strongSwan Team, 2022), • WireGuard (Hülsing et al., 2021). | • WireGuard operates independently and is not governed by standardization processes. It exclusively utilizes UDP traffic, necessitates specific infrastructure for operation, and raises privacy concerns as it mandates user data logging. In contrast, OpenVPN exhibits a longer initiation time compared to both StrongSwan and WireGuard. It offers enhanced compatibility but may have trade-offs, such as potentially lower throughput and longer ping times compared to WireGuard, which excels in circumventing censorship challenges (Dekker and Spaans, 2020). • Both OpenVPN and StrongSwan provide flexibility in authentication, supporting a range of methods including preshared keys and digital certificates for secret sharing. However, WireGuard exclusively supports preshared keys. • When generating post-quantum preshared keys from passwords, it is essential to be cautious about low entropy, a lack of full randomness, and vulnerability to attacks like Birthday attacks and Grover's algorithm. Moreover, preshared keys can be susceptible to side-channel attacks due to observable timing differences and cache access patterns, as indicated in a specific table (Table 1). • Post-quantum digital certificates come with drawbacks outlined in Table 11. These drawbacks should be considered when evaluating the use of digital certificates in a post-quantum security context. | • Info. Disclosure, • DoS | • **Enhancing Preshared Key Security:** (a) Enforce strong password policies for preshared keys, ensuring they have sufficient entropy to resist attacks like Birthday attacks and Grover's algorithm, (b) Implement monitoring and alert systems to detect unusual access patterns indicative of side-channel attacks on preshared keys. • **Detecting/Preventing Side-Channel Attacks:** To detect unusual access patterns indicative of side-channel attacks on preshared keys, implement monitoring and alert systems or use solutions mentioned in Table 1 to make preshared keys immune against side-channel attacks. • **Addressing Certificate-Based Drawbacks:** (a) Be aware of the drawbacks associated with post-quantum digital certificates, as outlined in Table 11, (b) Implement strategies like frequent certificate renewal and revocation to mitigate potential risks, Employ certificate management tools that support post-quantum algorithms and provide options for rapid updates. • **Optimizing Performance Trade-Offs:** (a) Carefully evaluate the specific use cases of OpenVPN, StrongSwan, and WireGuard, (b) Choose OpenVPN for compatibility, WireGuard for low latency and bypassing censorship, based on your priorities, (c) Optimize network infrastructure and use Quality of Service (QoS) mechanisms to mitigate potential performance trade-offs, (d) For WireGuard, consider its unique benefits in terms of performance, but also address privacy concerns by implementing data retention policies and encrypting logged data. |

The transition to quantum-resistant alternatives is not a matter of if, but when. It is crucial to ensure the confidentiality, integrity, and availability of data and communications in a post-quantum era. One of the fundamental components in this endeavor is the IKE protocol. IKE plays a pivotal role in establishing secure connections and managing encryption keys in IPsec-based VPNs. As quantum computing poses a considerable threat to IKE's DH key exchange, it is urgent to implement quantum-resistant IKE, as detailed in RFC 8784 titled "Mixing Preshared Keys in the IKE (v2) for Post-Quantum Security". This quantum-resistant IKE is a vital step towards securing communications in the quantum era. As quantum computing continues to advance, addressing quantum threats in the Internet Layer becomes increasingly critical. Organizations must carefully consider the transition to quantum-resistant alternatives to ensure the confidentiality, integrity, and availability of their data and communications in a post-quantum era. To address the challenges and vulnerabilities associated with OpenVPN, StrongSwan, and WireGuard in the context of adapting IPsec protocols with post-quantum cryptography, consider the following measures:

1. *WireGuard's Unique Operation and Privacy Concerns:* Organizations should evaluate their specific use cases. OpenVPN is suitable for compatibility, while WireGuard excels in low latency and bypassing censorship. To address performance trade-offs, optimize network infrastructure and use Quality of Service (QoS) mechanisms. Privacy concerns can be mitigated by implementing data retention policies and encrypting logged data.

2. *Authentication Flexibility:* Both OpenVPN and StrongSwan support multiple authentication methods. Enforce strong password policies for pre-shared keys and monitor for unusual access patterns that could indicate side-channel attacks on these keys.

3. *Vulnerability of Pre-shared Keys to Attacks:* Organizations should enforce strong password policies to ensure that pre-shared keys have sufficient entropy to resist attacks. Implement monitoring and alert systems to detect unusual access patterns indicative of side-channel attacks. Enhance security against side-channel attacks by considering solutions outlined in Table 1.

4. *Post-Quantum Digital Certificates:* To address the drawbacks associated with post-quantum digital certificates, organizations should be aware of potential risks. Implement strategies such as frequent certificate renewal and revocation to mitigate these risks. Additionally, use certificate management tools that support post-quantum algorithms and provide options for rapid updates to enhance security. By implementing these countermeasures, organizations can strengthen the security and resilience of their IPsec protocols in the face of quantum threats.

In conclusion, the evolution of quantum computing demands a proactive approach to secure Internet Layer protocols, ensuring that data transmission remains confidential and resilient against emerging threats. The integration of quantum-resistant alternatives and the implementation of robust security measures are pivotal in adapting to the quantum era and safeguarding sensitive data and communications.

### 4.4. Network interface layer

In the TCP/IP model, the Network Interface Layer is a critical component of network communication. Its primary objective is to ensure the dependable and efficient transmission of data across physical network connections, addressing emerging security concerns, especially in the age of quantum computing. The Internet l facilitates data exchange among various software applications, including web browsers, email clients, and file transfer programs. It ensures that data is presented in a format understandable by applications and manages communication between diverse software on various devices. Additionally, it defines the protocols and services that applications use to interact with lower layers of the network stack. Fundamentally, the Network Interface Layer oversees tasks such as physical addressing, local area network (LAN) delivery, and the transmission of raw data bits. Protocols operating at this layer encompass Ethernet for wired LANs and PPP (Point-to-Point Protocol) for point-to-point connections. Security protocols at the Network Interface Layer are pivotal in safeguarding network access, including technologies like Wi-Fi/WPA (Wireless Fidelity/Wi-Fi Protected Access) and DECT (Digital Enhanced Cordless Telecommunications), ensuring secure wireless communication and data protection across physical network connections.

Quantum computing introduces potential vulnerabilities that can impact data security, even at higher network layers. While security protocols, such as TLS and SSH are quantum-resistant, quantum computers have the capability to intercept data at the Network Interface Layer, posing a significant security risk. To combat this challenge, quantum-resistant physical layer security protocols are being developed for the Network Interface Layer, incorporating techniques such as quantum-safe encryption to maintain data confidentiality in the face of quantum threats.

In summary, the Network Interface Layer's role is of paramount importance in the quantum era, and the implementation of quantum-resistant techniques and protocols at this layer is essential to ensure the overall security of network communications amid evolving security challenges. In the following sections, we will delve into Wi-Fi/WPA and DECT as two key security protocols within the Network Interface Layer and explore potential quantum-resistant solutions for them.

#### 4.4.1. Wi-Fi/WPA protocol

Wi-Fi, specifically WPA (Wi-Fi Protected Access) version 3, plays a pivotal role in securing wireless networks by implementing a range of robust security mechanisms. It utilizes authentication, authenticated encryption, key derivation, and management frame protection to establish secure wireless networks (Soliman et al., 2009; Montville et al., 2009; Kelly et al., 2009; Dierks and Allen, 1999; Aboba and Wood, 2005; Krishnan and Kavanagh, 2011; Sakane et al., 2009). WPA's primary objectives are to ensure secure Wi-Fi connections and maintain the confidentiality, integrity, and authentication of transmitted data. However, as with many cryptographic protocols, WPA is susceptible to quantum computing threats.

Quantum computing introduces vulnerabilities to WPA's security. Notably, Shor's algorithm, a quantum algorithm, has the potential to compromise the underlying public-key cryptography used in WPA, including RSA and ECDH key exchange. This poses a risk to the confidentiality and integrity of Wi-Fi communications. Additionally, Grover's algorithm can weaken symmetric cryptography, such as AES and HMAC-SHA-3, which are integral to encryption and hashing in WPA. Consequently, quantum attacks could facilitate spoofing, tampering, and information disclosure against Wi-Fi networks.

In summary, while WPA version 3 enhances the security of Wi-Fi networks through advanced encryption and authentication mechanisms, it remains vulnerable to quantum computing threats. Quantum algorithms like Shor's and Grover's have the potential to undermine the cryptographic foundations of WPA, necessitating vigilance from network administrators and consideration of post-quantum cryptographic solutions as quantum computing technology advances.

To provide quantum-resistant WiFi/WPA and safeguard wireless networks against the impending threat posed by quantum computing, traditional RSA, ECDH, and ECDSA protocols should be replaced with post-quantum public-key cryptographic algorithms (Sakane et al., 2009). For symmetric cryptography, using longer keys suffices to provide an alternative post-quantum solution. Additionally, in the implementation of WiFi/WPA (v3) Enterprise version, TLS is used for secure communication, which should be replaced with possible post-quantum TLS, and PQ certificates in the enterprise version should be updated (Sakane et al., 2009). However, this migration is not without its challenges.

One major concern is the potential for substantial communication overhead due to the use of post-quantum encryption algorithms. This can lead to network congestion and fragmentation issues, triggering retransmissions, increasing latency, and potentially causing denial of service (DoS) attacks. Moreover, there is a risk of information disclosure through side-channel attacks, exploiting observable timing differences and cache access patterns.

To mitigate these challenges, several countermeasures can be adopted. Congestion can be controlled through the implementation of congestion control mechanisms. Fragmentation and retransmission issues can be addressed by optimizing message sizes for post-quantum protocols. To handle the increased communication overhead and prevent DoS attacks, infrastructure redundancy and DDoS resilience measures, such as firewalls and DDoS protection appliances, can be deployed. Additionally, configurations can be fine-tuned to withstand DDoS attacks. To protect against side-channel attacks, security solutions mentioned in Table 1 can be implemented. Furthermore, the weaknesses of TLS and digital certificates, which are crucial components of the enterprise version of WPA (v3), should be addressed following the recommendations outlined in Table 11.

In summary, quantum-resistant WiFi/WPA enhances the security of wireless networks against quantum threats but requires careful planning and implementation of countermeasures to address the associated challenges such as communication overhead, congestion, fragmentation, and the risk of DoS and information disclosure through side-channel attacks.

### 4.4.2. DECT protocol

DECT (Digital Enhanced Cordless Telecommunications) version 6.0 is a communication protocol primarily designed for cordless voice, fax, data, and multimedia communication. It is often used in wireless PBX systems and Wireless Local Area Networks (WLANs) to provide secure and reliable wireless communication (European Telecommunications Standards Institute (ETSI), 2022). DECT incorporates security mechanisms such as "Pairing", "Per Call Authentication", and "Encryption" as part of its standard protocol to ensure the security of wireless communications. These security features protect against eavesdropping and unauthorized access to DECT-based communications. However, DECT is not immune to quantum computing threats. Quantum computers, specifically those employing Grover's algorithm, pose a significant risk to DECT's security by potentially breaking the AES-based encryption used in DECT's DSC2, compromising the confidentiality of voice communications. As a result, DECT faces vulnerabilities related to spoofing, information disclosure, and privilege elevation when exposed to quantum attacks.

To address these threats, quantum-resistant DECT aims to secure DECT communication against quantum computing. One approach is to adopt longer and truly random symmetric keys, making DECT systems more resilient to quantum attacks, particularly against spoofing attempts. It is essential for DECT and similar protocols to continuously evolve their security measures in response to emerging quantum threats, including regular updates to cryptographic techniques and key management practices to stay ahead of potential vulnerabilities as quantum computing technology advances.

### 4.4.3. Final discussion: Adapting network interface layer security for the quantum era

In summary, the Network Interface Layer assumes a pivotal role in the realm of data transmission and security within the digital landscape. Its primary functions encompass the management of physical data transmission over network mediums and the assurance of data security.

Within the Network Interface Layer, protocols like WiFi/WPA serve as vital components responsible for both data transmission and security. However, it is imperative to recognize that these protocols are not impervious to the emerging threats posed by quantum computing.

Quantum algorithms, such as Shor's and Grover's, have the potential to compromise the security mechanisms that underpin these protocols. A detailed analysis of these protocols, their standardization, core purposes, recommended cipher suites, and quantum-related vulnerabilities are provided in Table 8.

Looking ahead, Table 9 presents prospective quantum-resistant alternatives for the Network Interface Layer protocols. It not only elucidates the challenges and vulnerabilities associated with the adoption of these alternatives but also sheds light on the potential threats that quantum-resistant solutions may encounter. Moreover, it proposes countermeasures to effectively mitigate these threats.

In conclusion, as quantum computing continues to evolve, addressing quantum threats to the Internet becomes critical to maintaining the security and reliability of network communications. The transition to quantum-resistant protocols and the careful implementation of countermeasures are fundamental in strengthening our defenses against the burgeoning realm of quantum risks. Organizations must remain vigilant and adapt their security measures to combat these threats effectively, as the security of data and network communications is of utmost importance in the digital era.

### 4.5. Quantum-safe certificates: Safeguarding certificates against quantum threats

X.509 certificates serve as a fundamental component of digital security, underpinning the foundation of secure digital communications. These certificates play a crucial role in supporting Public Key Infrastructure (PKI) and identity systems, ensuring the authenticity and integrity of digital entities in the digital world. They serve multiple vital functions, including data encryption, verifying the identities of communication partners, and maintaining data exchange confidentiality and integrity (Cooper et al., 2008a; IBM, 2022). These security features are achieved through a combination of cryptographic techniques, including public-key cryptography (e.g., RSA, ECDHE, ECDSA) and symmetric cryptography (such as AES and SHA2) (Cooper et al., 2008b). However, it is important to acknowledge that these cryptographic methods may face vulnerabilities in the future due to advances in quantum computing, including Shor's and Grover's algorithms (see Table 10).

In the context of quantum-safe network security protocols, quantum-resistant X.509 certificates have emerged as a crucial defense mechanism in digital security. These certificates build upon the foundations of X.509 certificates but are specifically designed to withstand the challenges posed by quantum computing. They ensure the longevity of cryptographic operations in a post-quantum environment by adopting post-quantum (PQ) cryptographic algorithms, as discussed in Section 3.2. Furthermore, to enhance the security of these certificates, they utilize symmetric cryptography with extended key lengths. This strategic combination of PQ public-key algorithms and more robust symmetric encryption keys provides an effective defense mechanism against potential threats from quantum computers. Quantum computers have the capability to compromise traditional encryption methods, making this adaptation essential for maintaining the integrity and security of X.509 certificates in the age of quantum computing (Refer to Table 10 for further details).

However, integrating PQ certificates is not without challenges and potential vulnerabilities. PQ certificates are generally larger in size, which can increase network traffic and susceptibility to Denial of Service (DoS) attacks due to resource depletion. The division of certificates and subsequent retransmissions can cause network congestion, potentially disrupting communication. Additionally, concerns arise about client certificate buffer overflows, posing a risk of code injection attacks by remote unauthenticated adversaries. Another challenge is the absence of a fallback mechanism in case of security or implementation issues with quantum-safe algorithms. Furthermore, PQ certificates may be vulnerable to data exfiltration through side-channel attacks on PQ algorithms.

**Table 8**
Network interface layer protocols, quantum threats and vulnerabilities.

| TCP/IP Layer | Protocols | Standard/Specification | Main components | Objectives | Recommended cipher suite | Quantum threats |
|---|---|---|---|---|---|---|
| Network Interface | WiFi/WPA (v3) | RFC 5416 (Soliman et al., 2009), RFC 5417 (Montville et al., 2009), RFC 5418 (Kelly et al., 2009), RFC 2246 (Dierks and Allen, 1999), RFC 4282 (Aboba and Wood, 2005), RFC 6418 (Krishnan and Kavanagh, 2011), RFC 5413 (Sakane et al., 2009) | • Authentication, • Encryption, • Key Derivation and Confirmation, • Robust Management Frame Protection. | • Enhancing security handshake via Wi-Fi DPP to establish secure wireless networks, • Supporting robust authentication methods such as EAP-TLS Enterprise to ensure the identity of network users, • Enabling data confidentiality through authenticated encryption methods, • Ensuring the integrity of transmitted data through the use of HMAC and secure hash algorithms. | • Public-Key Crypto: RSA, ECDH key exchange and ECDSA (broken by Shor's Algo.). • Symmetric Crypto: AES, HMAC-SHA-3, AES, GCMP, BIP (Weakened by Grover's Algo.) | • Spoofing, • Tempering, • Info. Disclosure. |
| | DECT (v6.0) | ETSI EN 300 175 (European Telecommunications Standards Institute (ETSI), 2022) | • Physical Layer, • Medium Access Control (MAC) Layer, • Link Control Layer, • Data Link Control (DLC) Layer. | • Enabling versatile communication services including cordless voice, fax, data, and multimedia communications, along with support for WLAN (Wireless Local Area Network) and wireless PBX (Private Branch Exchange), • Enhancing handset authentication using DSAA2, • Ensuring voice communication confidentiality by by encrypting the voice stream using DSC2, a security mechanism based on AES 128 encryption, • Providing secure connection authorization between handsets and base stations through subscription mechanisms. | • Public-Key Crypto: - • Symmetric Crypto: DSAA2 and DSC2 which are based on AES (Weakened by Grover's Algo.) | • Spoofing, • Info. Disclosure, • Elevation of Privilege. |

**Table 9**
Quantum-resistant network interface layer protocols: Safeguarding against quantum threats.

| TCP/IP Layer | Protocols | Possible quantum-resistant solutions | Challenges and vulnerabilities | Possible threats | Possible countermeasures |
|---|---|---|---|---|---|
| Network Interface | WiFi/WPA | • Wi-Fi/WPA (v3) can achieve quantum resistance through the following measures: (a) Transition from classic public-key cryptography used in Wi-Fi/WPA (such as RSA, ECDH, and ECDSA) to post-quantum (PQ) cryptography. (b) Enhance Symmetric Crypto by implementing longer key lengths, thereby fortifying the overall security of the Wi-Fi/WPA protocol. (c) Upgrade TLS to Quantum-Resistant TLS to ensure secure data transmission, preserving the integrity and confidentiality of communication within the enterprise. (d) Update PQ Certificates within the enterprise version, as outlined in Section 4.5, to ensure that the certificate infrastructure aligns with the overall quantum-resistant strategy of the Wi-Fi/WPA (v3) protocol and remains quantum-resistant. | • Huge communication overhead, • Network congestion, • Fragmentation issues triggering retransmission, • Possibility of Denial of Services, • Info. Disclosure via side-channel attacks (mentioned in Table 1) as a result of observable timing differences and cache access patterns. • Vulnerabilities related to TLS and digital certificate when the enterprise version of WPA (v3) is used (mentioned in Table 11 for certificates and in Section 4.2 for TLS). | • Info. Disclosure, • DoS | • **To control congestion:** Implement congestion control mechanisms to prevent or mitigate congestion issues effectively. This involves optimizing network traffic to prevent congestion and ensure smooth data flow. It is essential to monitor network usage and implement policies to manage congestion effectively (Tariq, 2023). • **To avoid fragmentation and trigger retransmission:** Address the increased message size resulting from quantum-resistant protocols by optimizing the message size. Ensure that the message size aligns with the capabilities of the network infrastructure to reduce fragmentation and the need for retransmissions (Yaacoub, 2023). • **To increase tolerance against communication overhead and prevent DoS:** Provide the necessary bandwidth to accommodate the communication overhead introduced by post-quantum encryption algorithms. Additionally, enhance network resilience by deploying DDoS protection solutions, such as firewalls and DDoS protection appliances. Configuration adjustments can also help withstand potential DoS attacks (Industrial Internet of Things (IIoT) Consortium, 2021). • **To immunize against side-channel attacks:** Implement security solutions outlined in Table 1 to protect against side-channel attacks. This includes addressing observable timing differences and cache access patterns to enhance security against these vulnerabilities (Zou, 2015). • **For TLS and digital certificate vulnerabilities:** Refer to the countermeasures provided in Table [tab:Post-migration-Cert] for certificates and other recommended solutions for TLS. This step is crucial to address vulnerabilities in TLS and digital certificates, particularly when using the enterprise version of WPA (v3) with post-quantum security measures (White House, 2022). |
| | DECT (v6) | • Utilizing DECT with extended symmetric key lengths. | • Inadequate Randomness in Shared Secrets: The reliance on the same PIN as a shared secret and a single random number for generating a symmetric key raises concerns about the randomness of these components. Such deficiencies could render the system vulnerable to attacks like the Birthday attack and Grover's algorithm (Grover, 2023). | • Spoofing | • **Enhancing Key Generation Security:** To mitigate the vulnerability stemming from inadequate randomness, it is crucial to provide a strongly random and sufficiently long symmetric key. This key should be designed to resist easy recovery, thus bolstering the security of the DECT system against quantum threats. |

**Table 10**
X.509 certificate, quantum threats and vulnerabilities.

| Certificate | Standard/ Specification | Fields of certificate | Objectives | Recommended cipher suite | Quantum threats |
|---|---|---|---|---|---|
| X.509 (v3) (Cooper et al., 2008a; IBM, 2022) | RFC 5280 (Cooper et al., 2008b) | • Version number, • Serial number, • Signature Algorithm Identifier, • Issuer name, • Validity period, • Subject Name, • Subject's public key information, • Issuer Unique ID, • Subject Unique ID, • Extensions. | • Managing identity and security in computer networking and online interactions. This includes securing email, communications, and digital signatures, ensuring the security and integrity of online communications and transactions, and establishing trust in the identities of the entities involved in these interactions. • Managing subject and issuer names, allowing for their potential reuse over time. This is essential in scenarios where certificates need to be updated or renewed without disrupting operations. • Supporting extensions to enable the inclusion of additional functionality and information in the certificate, ensuring adaptability to ensure compatibility with evolving security requirements and standards. • Providing a range of security features, including authentication (ensuring the identity of parties involved), integrity (safeguarding data from tampering), confidentiality (encrypting sensitive information), and repudiation (allowing parties to repudiate actions when necessary). | • Public-Key Crypto: RSA, ECDHE, ECDSA (broken by Shor's Algo.) • Symmetric Crypto: AES, SHA2 (Weakened by Grover's Algo.) | • Spoofing, • Tampering, • Repudiation • Info. Disclosure. |

**Table 11**
Quantum-resistant X.509 certificate: Safeguarding against quantum threats.

| Certificate | Possible quantum-resistant solutions | Challenges and vulnerabilities | Possible threats | Possible countermeasures |
|---|---|---|---|---|
| X.509 | A potential quantum-resistant solution for X.509 certificates involves replacing the traditional public-key cryptography with PQ cryptographic algorithms mentioned in Section 3.2. In addition, enhancing the security of the X.509 certificates can be achieved by using symmetric cryptography with longer keys. This combination of post-quantum public-key algorithms and stronger symmetric encryption keys helps safeguard X.509 certificates against potential threats posed by quantum computers, which have the potential to break traditional encryption methods. | • **Increased Certificate Size:** Using PQ cryptographic algorithms can result in larger certificate sizes. This may impact network efficiency and storage requirements.<br>• **Possibility of Denial of Service:** Adversaries could exploit the resource-intensive nature of PQ algorithms to launch DoS attacks against systems using quantum-resistant X.509 certificates.<br>• **Fragmentation and Retransmission:** Larger certificate sizes may lead to fragmentation, which can trigger retransmission and introduce inefficiencies in data transfer.<br>• **Network Congestion:** The increased computational load required for PQ cryptography might lead to network congestion during certificate validation and key exchange processes.<br>• **Possibility of Client Certificate Buffer Overflow:** Adversaries might attempt to exploit vulnerabilities in the implementation of quantum-resistant X.509 certificates, potentially leading to buffer overflow attacks and code injection (Homoliak et al., 2014; Madan et al., 2005; SONICWALL, 2022).<br>• **No Fall Back Plan:** Relying solely on quantum-safe algorithms may pose a risk if security or implementation issues arise in the future. A lack of fallback options could leave systems vulnerable.<br>• Data Exfiltration via Side-Channel Attacks: Side-channel attacks on PQ algorithms could be used to exfiltrate sensitive data. Careful protection against side-channel vulnerabilities (mentioned in Table 1) is necessary to maintain security. | • Info. Disclosure.<br>• DoS. | • **To prevent DoS Attacks:** (a) ensure the availability of the required bandwidth to handle the increased computational load introduced by quantum-resistant algorithms, (b) build redundancy into the infrastructure to distribute the load and withstand DDoS attacks, (c) deploy DDoS resilience hardware and software modules such as firewalls, (d) consider adopting DDoS protection appliances to mitigate the impact of denial of service attacks, and (e) configure network hardware to detect and counter DDoS attacks (Garcia and Blandon, 2022a; Liu et al., 2018).<br>• **To avoid Fragmentation and Trigger Retransmission:** (a) take into account the larger size of quantum-safe certificates compared to traditional ones when designing network communication protocol, (b) ensure that network systems can handle the extreme size of messages when quantum-safe certificates are used (Müller et al., 2020; Beernink, 2022).<br>• **To Control Network Congestion:** (a) implement congestion control mechanisms to prevent or alleviate network congestion, especially during certificate validation and key exchange processes, and (b) monitor network traffic and dynamically adjust resources to handle congestion (Jay et al., 2018; Bohloulzadeh and Rajaei, 2020; Jay et al., 2019).<br>• **To Be Immune Against Stack-Based Buffer Overflow Attacks for Certificates on Servers by Malicious Clients:** (a) implement security measures such as Canary values to detect buffer overflows, (b) employ Data Execution Prevention (DEP) and Address Space Layout Randomization (ASLR) to prevent malicious code injection, (c) and continuously update and patch server software to address vulnerabilities that could be exploited by buffer overflow attacks (Zhou et al., 2022; Nicula and Zota, 2019).<br>• **To Avoid a Lack of Fallback:** (a) adopt crypto-agility, which involves supporting multiple quantum-safe algorithms, (b) develop the ability to switch between algorithms in the event of security or implementation issues with a particular quantum-safe algorithm, (c) stay informed about emerging post-quantum cryptography developments to ensure a secure transition (Ma et al., 2021; Ott et al., 2019; Ma, 2021). |

To mitigate these challenges and vulnerabilities, various countermeasures have been proposed. Organizations can ensure adequate network bandwidth to thwart DoS attacks and deploy DDoS resilience hardware/software modules like firewalls and DDoS protection appliances. Addressing fragmentation and retransmission issues requires careful consideration of message size when using quantum-safe certificates. Network congestion can be managed through the implementation of congestion control mechanisms. To minimize the risk of stack-based buffer overflow vulnerabilities in server certificates, security measures like Canary (Zhou et al., 2022), Data Execution Prevention (DEP) (Microsoft, 2023), and Address Space Layout Randomization (ASLR) (Nicula and Zota, 2019) can be employed. To address the absence of a fallback mechanism, organizations can adopt crypto-agility by supporting multiple quantum-safe algorithms.

Concerning cipher suites, quantum-safe X.509 certificates recommend the use of algorithms specified in Table 1, ensuring information security and strengthening defenses against potential DoS attacks. While the adoption of PQ certificates introduces new challenges, careful planning and the implementation of robust countermeasures can pave the way for a secure and quantum-resistant future in the realm of digital certificates and online security.

## 5. Hybrid approaches to ensure business continuity in quantum-safe network security protocols

In the realm of network security, ensuring business continuity is paramount. To achieve this, innovative strategies have emerged, leveraging classical and post-quantum cryptographic algorithms within protocols. These strategies aim to preserve backward compatibility while mitigating potential quantum computing threats (Florence, 2023). There are two main approaches in this context:

- **Non-Composite Protocol Approach:** The non-composite protocol approach is designed to enable a seamless transition to quantum-safe protocols. However, this method requires significant modifications, including adjustments to protocol fields, message flow, and libraries to support both classical and post-quantum protocols. Despite the need for extensive changes, it offers a pathway to enhanced security.
  Notable advantages of this approach include only a few changes of standards and applications/devices for the non-composite protocol, a smooth transition to a quantum-safe protocol, avoidance of fragmentation issues for quantum-safe protocols, the format of cryptographic elements related to the protocol remains unchanged, and minimal modifications to cryptographic elements and libraries for the protocol. However, there are some disadvantages, including the need for primary modifications in protocol fields, the message flow, or both to support both classical and post-quantum protocols, redundant information transmission, and required changes in protocol libraries. Quantum computing threats in this approach may include spoofing, tampering, repudiation, information disclosure, denial of service, and elevation of privileges when neither of the protocols used is secure against them.
- **Composite Protocol Approach:** In contrast, the composite protocol approach minimizes alterations to the core structure of protocols. It allows multiple component algorithms to run in parallel, maintaining the same protocol fields and message flow as single-algorithm versions. While this approach preserves the integrity of the protocol's structure, it does mandate adjustments to cryptographic elements.
  Key points related to this approach are protocol fields and message flow remain unchanged, and minimal changes are likely to be made to cryptographic libraries. The disadvantages involve required changes primarily in the formats of the cryptographic elements of the protocol, and most changes are likely to be made to cryptographic libraries. Quantum computing threats in this approach are similar to the non-composite protocol approach.

Both of these strategies must be implemented with careful consideration of potential quantum computing threats, which encompass spoofing, tampering, repudiation, information disclosure, denial of service, and elevation of privileges. These threats underscore the importance of continuously evaluating and updating security measures. Adapting to emerging vulnerabilities in network security is essential for safeguarding sensitive data and ensuring the resilience of network systems.

In summary, these strategies are pivotal in achieving business continuity in network security. They empower organizations to navigate the evolving landscape of quantum computing threats while maintaining the integrity of their protocols. A comprehensive summary of both strategies, including their advantages and disadvantages, as well as potential quantum computing threats, can be found in Table 12. Additionally, it is important to note that the mechanisms for creating hybrid protocols can be applied not only to entire protocols but also to sub-protocols, combining them for various protocol components, including negotiation, key exchange, or authentication.

### 5.1. Hybrid approach certificate: Ensuring business continuity in network security

Continuing from the previous section, the hybrid strategy for certificates is an essential component of the broader effort to ensure network security and business continuity in the face of potential quantum computing threats. This strategy seamlessly combines classical and post-quantum cryptographic solutions, acting as a bridge between quantum-safe and non-quantum-safe cryptographic (Bindel et al., 2017; Vogt and Funke, 2021; Bindel et al., 2019).

A hybrid strategy combines classical and post-quantum solutions to ensure backward compatibility and serves as a bridge between quantum-safe and non-quantum-safe cryptographic states. It ensures seamless connectivity during the migration period until the complete transition to a quantum-safe cryptographic state. In the hybrid strategy for certificates, multiple certificates (e.g., one classical and one post-quantum) are combined into a classical X.509 certificate, providing a flexible solution for servers to accommodate clients with different cryptography capabilities. The hybrid approach remains secure as long as at least one certificate in the combination remains unbroken. In this context, we surveyed existing hybrid approaches for certificates, their mechanisms, highlighting their advantages, disadvantages, and potential quantum computing (QC) threats in Table 13 while ensuring backward compatibility.

As part of the hybrid approach for protocols, a composite certificate is required. According to RFC 5280 (Cooper et al., 2008b), an X.509 certificate can only have one TBS certificate, meaning it can contain only one subject public key and is signed using only one CA signature. Creating a hybrid certificate that combines classical and post-quantum solutions is challenging. To overcome these challenges, different approaches have been proposed. These approaches aim to preserve backward compatibility while supporting both classical and quantum certificates.

The first approach, non-composite certificates, creates two distinct certificates: one for conventional algorithms and the other for post-quantum algorithms. The second approach (composite certificate) uses an extension mechanism to create a hybrid certificate. The extension field is used in two ways: (i) a post-quantum certificate is created first and then set as the extension in the classical certificate. However, this approach introduces information redundancy, increasing certificate sizes when the same information is repeated in both certificates. This extension mechanism requires changes in standards (e.g., RFC 5280) to store and verify two signatures and two public keys in a certificate; (ii) only extra information, such as PQ public key and signature information, is embedded into the extension to avoid redundancy of similar information in the certificate, facilitating a smooth transition to quantum-safe certificates. Nevertheless, this extension mechanism also necessitates changes in standards (e.g., RFC 5280) to store and verify two signatures and two public keys in a certificate.

**Table 12**

Hybrid approach for protocols: Ensuring business continuity in network security.

| Strategies | Mechanisms | Advantages | Disadvantages | Quantum threats |
|---|---|---|---|---|
| Non-composite Protocol (Florence, 2023) | Two separate protocols with separate certificates, maintaining component cryptographic element formats consistent with single-algorithm schemes. | • Only a few changes of standards and applications/devices for non-composite protocol. • Smooth transition to quantum-safe protocol. • Avoid fragmentation issues for quantum-safe protocol. • Format of cryptographic elements related to the protocol remains unchanged. • Cryptographic elements and libraries have minimal modifications for the protocol. | • Some primary modifications are required in protocol fields, the message flow, or both to support both classic and PQ protocols. • Redundant information is required to transmit in the non-composite protocol. • To implement the non-composite protocol, some modifications are required in protocol libraries. Also, some changes are needed to support parallel hierarchies. | • Spoofing, Tampering, Repudiation, Information Disclosure, DoS, and Elevation of Privileges (when none of the protocols used in the hybrid approach are secure against them). |
| Composite Protocol (Florence, 2023) | Composite hybrid protocols, maintaining protocol fields and message flow consistent with classic versions using single-algorithm schemes. | • Protocol fields and message flow remain unchanged. • Minimal changes are likely to be made to the cryptographic libraries. | • Required changes are primarily made to the formats of the cryptographic elements of the protocol. • To implement a composite protocol, most changes are likely to be made to the cryptographic libraries. | • Spoofing, Tampering, Repudiation, Information Disclosure, DoS, and Elevation of Privileges (when none of the protocols used in the hybrid approach are secure against them). |

**Table 13**

Hybrid approach for certificate: Ensuring business continuity in network security.

| Strategies | Mechanisms | Advantages | Disadvantages | Quantum threats |
|---|---|---|---|---|
| Non-composite Certificate (Bindel et al., 2017; Vogt and Funke, 2021; Bindel et al., 2019) | Two Separate Certificates | • Only few changes of standards and applications/devices for PQ one (Vogt and Funke, 2021), • Only moderate increase of certificate size for PQ one, • Smooth transition to quantum-safe certificates. | • Since the subject, issuer, and other metadata are repeated in both, non-composite certificates are slightly bigger than one, • PKI software needs to be changed to manage parallel hierarchies. | • Spoofing, Tampering, Repudiation, Info. Disclosure, DoS, and Elevation of Privileges (when none of the protocols used in the hybrid approach are secure against them.) |
| Composite Certificate (Bindel et al., 2017; Vogt and Funke, 2021; Bindel et al., 2019) | Post-quantum Cert. in Extension | • Combines security of pre- and post-quantum algorithms. | • Since the subject, issuer, and other metadata are repeated in both original and in-extension certificates are slightly bigger than one, • Abrupt migration for all applications at the same time, • Needs changes of standards (e.g. RFC 5280 (Cooper et al., 2008a)) for two signatures and two public keys in a certificate, • Size of certificates increases the most. | • Spoofing, Tampering, Repudiation, Info. Disclosure, DoS, and Elevation of Privileges (when none of the protocols used in the hybrid approach are secure against them.) |
| | PQ Public key and Signature Information in the Extension | • Smooth transition to quantum-safe certificates, • Combines security of pre- and post-quantum algorithms. | • Needs changes of standards (e.g. RFC 5280 (Cooper et al., 2008a)) to store and verify two signatures and two public keys in a certificate — Size of certificates increases | • Spoofing, Tampering, Repudiation, Info. Disclosure, DoS, and Elevation of Privileges (when none of the protocols used in the hybrid approach are secure against them.) |

# 6. Conclusions

In this comprehensive study, we have undertaken an extensive exploration of quantum computing and its profound ramifications within networked environments. The immense processing capabilities of quantum computing have cast a formidable shadow over conventional cryptographic techniques, necessitating a thorough examination of quantum-safe security measures.

Our journey began with a meticulous examination of quantum algorithms, highlighting their transformative potential in the domain of computing security. The insights gleaned from this endeavor underscore the imminent threat quantum computing poses to traditional encryption methodologies, illuminating vulnerabilities deeply rooted in conventional network security. This research has successfully accentuated the pressing need for quantum-safe security protocols. We conducted an exhaustive vulnerability assessment, rigorously evaluating various network protocols for their resilience against simulated quantum attacks. Through controlled experiments, leveraging libraries like LibOQS via Python language-specific wrapper (Open Quantum Safe (OQS), 2024), we evaluated critical parameters, including encryption/decryption overhead and resource utilization. Our findings not only underscore the superiority of quantum-safe protocols but also highlight their ability to transcend the limitations of classical cryptographic techniques. The empirical evidence unequivocally establishes the superior resistance of quantum-safe protocols to quantum attacks, safeguarding data confidentiality and integrity in quantum computing environments while imposing minimal resource overhead. This empirical strength positions quantum-safe protocols as a critical component in securing networked environments against quantum threats.

In recognizing the strengths of our approach, it is equally important to acknowledge its limitations. Key distribution mechanisms and computational performance challenges warrant careful consideration, particularly in resource-constrained settings. These aspects must be addressed as we transition to quantum-safe security measures. Our approach's distinctiveness becomes even more apparent when viewed in the context of a comparative analysis. Traditional cryptographic protocols are increasingly susceptible to quantum attacks, while alternative quantum-resistant approaches often necessitate extensive infrastructure overhauls. In contrast, our hybrid approach offers a transitional solution that harmonizes the strengths of classical and quantum-safe protocols, providing effective network security without disruptive alterations.

As quantum computing technology continues to advance and the quantum threat landscape evolves, we emphasize the need for ongoing research and development. Future avenues for exploration include the optimization of quantum-safe protocols for enhanced performance and efficiency, innovative strategies for quantum-resistant key management, and vigilant monitoring of quantum advancements and their network security implications.

In conclusion, this scientific survey reinforces the critical importance of quantum-safe network security in light of evolving threats. The empirical results, along with practical case studies, substantiate the viability of our approach. In an era where quantum computing has ceased to be a distant future and is rapidly becoming a reality, the imperative for proactive action is clear. Our approach serves as the vanguard for a secure transition into the quantum age, safeguarding networked environments with an unparalleled level of resilience and adaptability.

## Declaration of competing interest

The authors declare that they have no known competing financial interests or personal relationships that could have appeared to influence the work reported in this paper.

## Data availability

No data was used for the research described in the article.

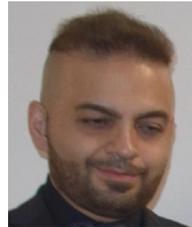

**Yaser Baseri**, PhD, is a cybersecurity researcher at Université de Montréal. Previously, he led the research and development team at the Canadian Institute for Cybersecurity. Dr. Baseri holds a Ph.D. in Computer Science and gained experience at CIRRELT. His research focuses on post-quantum cryptography, privacy-preserving technologies, cyber threat analytics, and risk assessment. This expertise, coupled with his industry partnerships (Scotiabank, RBC, Bell, Siemens, and the Canadian Internet Registry Authority), positions him as a respected researcher actively contributing to advancements in cybersecurity.

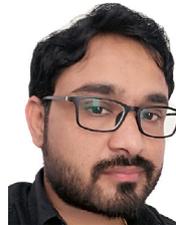

**Vikas Chouhan** is a postdoctoral fellow at the Canadian Institute for Cybersecurity (CIC) at the University of New Brunswick. He earned his M.Tech and Ph.D. in Computer Science and Engineering from the Indian Institute of Technology, Roorkee. An author with a significant portfolio of research publications, Dr. Chouhan has participated on program committees of various international conferences and served as a reviewer. His research interests are diverse, spanning cloud security, blockchain technology, and the security challenges posed by quantum computing.

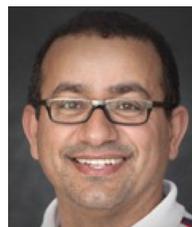

**Abdelhakim Hafid** holds a full professorship at the University of Montreal and is the founding director of the Network Research Lab and the Montreal Blockchain Lab. Also a Research Fellow at CIRRELT, he has a rich publication record with over 240 papers and three US patents to his name. His industry experience is extensive, having worked at Bellcore and held leadership roles in several telecommunications startups. Dr. Hafid's research interests are broad, including IoT, fog/edge computing, blockchain technology, and intelligent transport systems.